\documentclass[preprint, times]{aastex63}
\pdfoutput=1

\usepackage{booktabs, gensymb}
\usepackage{amsmath}
\usepackage{rotating}
\usepackage{subfigure}
\newcommand{\NA}{---}
\accepted{\today}
\submitjournal{ApJ}
\shorttitle{Reconstructing EUV of Cool Dwarfs}
\shortauthors{Duvvuri et al.}
\graphicspath{{./}{figures/}}
\begin{document}
\title{Reconstructing the Extreme Ultraviolet Emission of Cool Dwarfs Using Differential Emission Measure Polynomials}
\correspondingauthor{Girish M. Duvvuri}
\email{girish.duvvuri@gmail.com}
\author[0000-0002-7119-2543]{Girish M. Duvvuri}
\affiliation{Department of Astrophysical and Planetary Sciences, University of Colorado, Boulder, CO 80309, USA}
\affiliation{Laboratory for Atmospheric and Space Physics, University of Colorado, 600 UCB, Boulder, CO 80309}
\affiliation{Center for Astrophysics and Space Astronomy, University of Colorado, 593 UCB, Boulder, CO 80309}
\author[0000-0002-4489-0135]{J. Sebastian Pineda}
\affil{Department of Astrophysical and Planetary Sciences, University of Colorado, Boulder, CO 80309, USA}
\affiliation{Laboratory for Atmospheric and Space Physics, University of Colorado, 600 UCB, Boulder, CO 80309}
\author[0000-0002-3321-4924]{Zachory K. Berta-Thompson}
\affil{Department of Astrophysical and Planetary Sciences, University of Colorado, Boulder, CO 80309, USA}
\affiliation{Center for Astrophysics and Space Astronomy, University of Colorado, 593 UCB, Boulder, CO 80309}
\author[0000-0003-2631-3905]{Alexander Brown}
\affiliation{Center for Astrophysics and Space Astronomy, University of Colorado, 593 UCB, Boulder, CO 80309}
\author[0000-0002-1002-3674]{Kevin France}
\affiliation{Department of Astrophysical and Planetary Sciences, University of Colorado, Boulder, CO 80309, USA}
\affiliation{Laboratory for Atmospheric and Space Physics, University of Colorado, 600 UCB, Boulder, CO 80309}
\affiliation{Center for Astrophysics and Space Astronomy, University of Colorado, 593 UCB, Boulder, CO 80309}
\author[0000-0001-7458-1176]{Adam F. Kowalski}
\affiliation{Department of Astrophysical and Planetary Sciences, University of Colorado, Boulder, CO 80309, USA}
\affiliation{Laboratory for Atmospheric and Space Physics, University of Colorado, 600 UCB, Boulder, CO 80309}
\affiliation{National Solar Observatory, University of Colorado Boulder, 3665 Discovery Drive, Boulder, CO 80303, USA.}
\author[0000-0003-3786-3486]{Seth Redfield}
\affil{Astronomy Department and Van Vleck Observatory, Wesleyan University, Middletown, CT 06459, USA}
\author[0000-0001-9361-6629]{Dennis Tilipman}
\affil{Department of Astrophysical and Planetary Sciences, University of Colorado, Boulder, CO 80309, USA}
\affiliation{National Solar Observatory, University of Colorado Boulder, 3665 Discovery Drive, Boulder, CO 80303, USA.}
\author{Mariela C. Vieytes}
\affil{Instituto de Astronomía y Física del Espacio, CC 67 Suc. 28, 1428 Buenos Aires, Argentina}
\author[0000-0001-9667-9449]{David J. Wilson}
\affil{McDonald Observatory, University of Texas at Austin, Austin, TX 78712}
\author[0000-0002-1176-3391]{Allison Youngblood}
\affiliation{Laboratory for Atmospheric and Space Physics, University of Colorado, 600 UCB, Boulder, CO 80309}
\author[0000-0001-8499-2892]{Cynthia S. Froning}
\affil{McDonald Observatory, University of Texas at Austin, Austin, TX 78712, USA}
\author[0000-0003-4446-3181]{Jeffrey Linsky}
\affil{JILA, 440 UCB, University of Colorado, Boulder, CO 80309}
\author[0000-0001-5646-6668]{R. O. Parke Loyd}
\affil{School of Earth and Space Exploration, Arizona State University, Tempe, AZ 85287, USA}
\author{Pablo Mauas}
\affil{Instituto de Astronomía y Física del Espacio, CC 67 Suc. 28, 1428 Buenos Aires, Argentina}
\affil{Dpto. de Fisica, Facultad de Ciencias Exactas y Naturales (FCEN), Universidad de Buenos Aires (UBA), Buenos Aires, Argentina}
\author{Yamila Miguel}
\affil{Leiden Observatory, P.O. Box 9500, 2300 RA Leiden, The Netherlands}
\author[0000-0003-4150-841X]{Elisabeth R. Newton}
\affil{Department of Physics and Astronomy, Dartmouth College, Hanover, NH 03755, USA}
\author[0000-0003-1620-7658]{Sarah Rugheimer}
\affil{University Oxford, Atmospheric, Oceanic, and Planetary Physics Department, Clarendon Laboratory, Sherrington Road, Oxford OX1 3PU, UK}
\author{P. Christian Schneider}
\affil{Hamburder Sternwarte, Gojenbergsweg 112, D-21029 Hamburg, Germany}
\begin{abstract}
Characterizing the atmospheres of planets orbiting M dwarfs requires understanding the spectral energy distributions of M dwarfs over planetary lifetimes. Surveys like MUSCLES, HAZMAT, and FUMES have collected multiwavelength spectra across the spectral type's range of $T_{\textrm{eff}}$ and activity, but the extreme ultraviolet flux (EUV, 100 to 912 $\textrm{\AA}$) of most of these stars remains unobserved because of obscuration by the interstellar medium compounded with limited detector sensitivity. While targets with observable EUV flux exist, there is no currently operational facility observing between $150$ and $912\,\textrm{\AA}$. Inferring the spectra of exoplanet hosts in this regime is critical to studying the evolution of planetary atmospheres because the EUV heats the top of the thermosphere and drives atmospheric escape. This paper presents our implementation of the differential emission measure technique to reconstruct the EUV spectra of cool dwarfs. We characterize our method's accuracy and precision by applying it to the Sun and AU Mic. We then apply it to three fainter M dwarfs: GJ 832, Barnard's Star, and TRAPPIST-1. We demonstrate that with the strongest far ultraviolet (FUV, 912 to 1700 $\textrm{\AA}$) emission lines, observed with \emph{Hubble Space Telescope} and/or \emph{Far Ultraviolet Spectroscopic Explorer}, and a coarse X-ray spectrum from either \emph{Chandra X-ray Observatory} or \emph{XMM-Newton}, we can reconstruct the Sun's EUV spectrum to within a factor of 1.8, with our model's formal uncertainties encompassing the data. We report the integrated EUV flux of our M dwarf sample with uncertainties between a factor of 2 to 7 depending on available data quality.
\end{abstract}
\keywords{}
\section{Introduction} \label{sec:intro}
The discovery and characterization of exoplanets has been accompanied by an increased interest in the properties of M dwarf stars as potential hosts for habitable planets. M dwarf planetary systems are abundant, not only because M dwarfs are $\gtrsim 70 \%$ of all stars in the Milky Way \citep{Henry2006, Winters2015}, but also because M dwarfs also have an intrinsically high planet occurrence rate compared to their hotter and more massive siblings \citep{Dressing2015}. Moreover, these systems' physical properties benefit their detection and characterization: once for their abundance, twice for the large transit depths of terrestrial planets projected against small stellar radii, and thrice for the short orbital periods of planets with Earth-comparable instellation. The \emph{Transiting Exoplanet Survey Satellite} (\emph{TESS}) is predicted to find 1300 planets orbiting M dwarfs \citep{Ballard2019}, roughly 10 of which will be terrestrial worlds suitable for atmospheric characterization with the \emph{James Webb Space Telescope} \citep{Barclay2018}.

These M dwarf planetary systems provide a useful sample for study, but some caution that the ``habitability" potential of these systems may be poor or non-existent \citep{Scalo2007, Shields2016}. An M dwarf is a tempestuous host, prone to flaring \citep{Hawley1993, Kowalski2009, Loyd2018a, Loyd2018b}, particularly when young, where a mid-to-late M dwarf's definition of ``young" lasts for billions of years \citep{West2008}. Compared to the Sun, M dwarfs emit a much higher fraction of their bolometric flux in the ultraviolet regime \citep{West2004, Jones2016}. The extreme ultraviolet region (EUV, defined here as 100 - 912 $\textrm{\AA}$) is particularly responsible for heating and ionizing the upper atmosphere of planets, dumping energy into the system and potentially driving atmospheric escape \citep[e.g.][]{Sekiya1980, SanzForcada2010, OwenJackson2012, TianIda2015, ZahleCatling2017}. Any attempt to study an exoplanet atmosphere's evolution must be informed by the radiation field it is subject to over the entirety of its lifetime \citep{PenzMicela2008, Claire2012, Peacock2020}. But directly measuring the EUV flux is impeded by the same mechanism that makes it important for planet atmospheres: its interactions with atomic hydrogen and helium mean that the interstellar medium blocks some of the flux from this spectral region for most stars \citep{CoxReynolds1987, France2019}. This problem is exacerbated for M dwarfs since the closest M dwarfs with observable EUV flux either have noisy data from the \emph{Extreme Ultraviolet Explorer} (\emph{EUVE}) or no data at all \citep{Craig1997, Linsky2014, France2016}, and there is no presently available dedicated EUV observatory to remedy the situation \citep{France2019}.

In the absence of direct observation, we must turn to theoretical inference. \citet{Peacock2019a} and \citet{Peacock2019b} use the \texttt{PHOENIX} 1D stellar atmosphere code \citep{Hauschildt1993, HauschildtBaron2006, BaronHauschildt2007} to model the non-LTE radiative transfer through the chromospheres and transition regions of M dwarfs but do not include a corona. \citet{Fontenla2016} adjusts the temperature and pressure profiles of a 1D stellar atmosphere until the model agrees with the available spectral data, but this takes time to do well and has to be specific to each star. These semi-empirical methods require quasi-simultaneous observations from optical to X-ray wavelengths. All known atomic and molecular processes and species have to be taken into account in each layer, solving the NLTE coupled system to match observations of many lines and continua across the spectral range. This requires a reliable atomic database and laborious fine-tuning to be successful.

Taking a more empirical approach, \citet{Linsky2014} identifies correlations between Lyman-$\alpha$ and EUV flux, while \cite{Youngblood2017} identifies correlations between far-ultraviolet (FUV, 912 to 1700 $\textrm{\AA}$) lines and the Lyman-$\alpha$ flux, chaining these correlations to the \citet{Linsky2014} relations to predict the EUV flux in turn. A drawback of this method is that the uncertainty on each correlation introduces scatter into the predicted EUV flux while the sample is insufficiently large to investigate the effects of both effective temperature $T_{\textrm{eff}}$ and stellar activity. \citet{France2018} correlates certain FUV lines with the EUV flux between 90 and 360 $\textrm{\AA}$ directly, leading to much less scatter in the predicted flux and accounting for both $T_{\textrm{eff}}$ and stellar activity in their sample, but this still leaves us with $\sim 600$ $\textrm{\AA}$ of EUV flux to estimate.

These limitations of existing methods lead us to use the differential emission measure (DEM), a technique for EUV spectral synthesis adapted from an earlier technique called the emission measure distribution. \citet{Pottasch1963} defined the emission measure distribution $\equiv \frac{n_{\textrm{O}}}{n_{\textrm{H}}}\int n_e^2 \, ds$ as the integral of the electron number density squared ($n_e^2$) along the line of sight $s$ weighted by the relative abundance of oxygen to hydrogen ($\frac{n_\textrm{O}}{n_\textrm{H}}$), to describe the plasma environment of the upper layers of the Sun's atmosphere. This assumed that the Sun's upper atmosphere could be approximated as a series of spherical shells of increasing temperature, and all emission lines were produced by collisional excitation and spontaneous radiative decay within restricted spatial regions.

As this picture of spherical symmetry broke down, the differential emission measure was developed to keep the same 1-dimensional simplification to temperature but account for the spatial ambiguity of a photon's origin (see \citealt{Mariska1992} for a detailed overview of the method's history). The differential emission measure uses a similar integral expression over a limited temperature range to estimate the density and temperature environment of ions emitting an observed line, allowing one to then use those environmental conditions to estimate the flux from emission lines that cannot be observed but should be emitted by the same parcel of plasma. When UV detector technology was in its infancy and instruments had poor flux calibration,  differing by factors of $\gtrsim 2$ in different wavelength regimes, the DEM could be used to estimate the subset of solar emission line fluxes with poor data from other lines that were thought to have more accurate and precise data \citep{Warren1998}. While the state of solar EUV data has improved, the opacity of the interstellar medium and low sensitivities of previous and current EUV-capable observatories present a similar spectral synthesis problem for distant stars. Variations of the DEM have been applied to other stars like AU Mic by \citet{Pagano2000}, $\alpha$ Centauri A and B by \cite{Ayres2014}, and HD 209458 by \citet{Louden2017} to infer the EUV flux from these stars. \citet{SanzForcada2011} developed scaling relations between X-ray and EUV fluxes by applying the DEM method to a large sample of stars, but the paper's sample had few M dwarfs and lacked enough UV data to constrain the lower temperature end of the DEM for most of their stars.

In this paper we characterize our uncertainties in fitting the DEM and propagate them to our predictions of the EUV flux from M dwarfs. Our physical assumptions and setup are similar to the method described and used by \citet{Warren1998} to model the EUV irradiance of the Sun, described in Section \S\ref{sec:dem}. The specifics of our implementation are described in Section \S\ref{sec:implementation} and we test our method against data from the Sun in Section \S\ref{sec:sun}. In Section \S\ref{sec:au_mic} we apply our method to AU Mic, a $\sim 10 - 20$ Myr old M1 star at a distance of 9.979 pc \citep{MacGregor2013, Plavchan2020}. We compare our DEMs of the Sun and AU Mic to previous DEMs published in the literature and available in the \texttt{CHIANTI} atomic database \citep{Dere1997, DelZanna2015} in Section \S\ref{sec:literature_comparison}.

We compare our predicted spectra for the Sun and AU Mic to data in detail in Section \S\ref{sec:final_model_spectra} and in Section \S\ref{sec:case_studies} we apply our method to different case studies:  GJ 832, a planet-hosting M2 V that has predicted EUV fluxes from \citet{Linsky2014} and semi-empirical models from both \citet{Fontenla2016} and \citet{Peacock2019b}; Barnard's Star, a $\sim 10$ Gyr old M4 with a candidate planet \citep{Ribas2018}, with contemporaneous X-ray and FUV data during quiescence and a flare \citep{France2020}; and TRAPPIST-1, an ultracool dwarf which hosts at least seven planets \citep{Gillon2017} and tests our ability to fit the DEM in an extremely low S/N regime (Wilson et al. submitted ). Our work shows that with \emph{Hubble Space Telescope} (\emph{Hubble} or \emph{HST}) measurements of a few FUV emission line fluxes and a coarse X-ray spectrum from \emph{Chandra} or \emph{XMM-Newton}, we can estimate the EUV spectrum with meaningful uncertainties for any star whose EUV flux is dominated by emission lines from the optically thin regions of the star's upper atmosphere.

\section{Differential Emission Measure}\label{sec:dem}
The following description of the DEM is adapted from \citet{Warren1998}. Many other formulations of the DEM and similar techniques exist, and  \citet{Mariska1992} explains them in more detail. Given an optically thin plasma in a collisionally dominated time-independent equilibrium with negligible collisional de-excitation, the radiance of a wavelength transition is given by

\begin{align}\label{eq:radiance}
    I_{ul} &= \frac{1}{4\pi} \int_{\mathrm{line-of-sight}} n_u A_{ul}  \frac{hc}{\lambda_{ul}} \, ds \; [\mathrm{erg} \, \mathrm{s}^{-1} \, \mathrm{cm}^{-2} \, \mathrm{sr}^{-1}],
\end{align}
\noindent
where $ul$ signifies a transition from an upper state $u$ to a lower state $l$, $A_{ul}$ is the Einstein rate coefficient of the transition, $\lambda_{ul}$ is the wavelength of the transition, $h$ is Planck's constant, and $c$ is the speed of light in a vacuum. This quantity is not a spectral density because it captures all of the emission from the spontaneous radiative decay without describing a line profile. We can rewrite this integral as

\begin{align}\label{eq:radiance_dem}
    I_{ul} &= \int_T G_{ul}(T) \cdot \Psi (T)\, dT \; [\mathrm{erg} \, \mathrm{s}^{-1} \, \mathrm{cm}^{-2} \, \mathrm{sr}^{-1}],
\end{align}
\noindent
where

\begin{align}\label{eq:gofnt}
    G_{ul}(T) &= \frac{n_u}{n_{\mathrm{ion}}} \frac{n_{\mathrm{ion}}}{n_{\mathrm{element}}}\frac{n_{\mathrm{element}}}{n_{\mathrm{H}}} \frac{1}{n_e} \frac{A_{ul}hc}{4\pi \lambda_{ul}} \; [\mathrm{erg} \, \mathrm{s}^{-1} \, \mathrm{cm}^{3} \, \mathrm{sr}^{-1}]
\end{align}
\noindent
is the transition's emissivity contribution function and the differential emission measure is

\begin{equation}\label{eq:dem}
    \Psi (T) = n_e n_{\mathrm{H}} \frac{ds}{dT} \; [\mathrm{cm}^{-5} \, \mathrm{K}^{-1}].
\end{equation}
$G_{ul}(T)$, the emissivity contribution function, describes the volume integrated power of a parcel of gas as a function of temperature. The function can be computed with a few ingredients: a stellar abundance to give us the ratio of the number density of any particular element's atoms to the number density of hydrogen atoms $\frac{n_{\mathrm{element}}}{n_{\mathrm{H}}}$, the assumption of collisionally dominated equilibrium (i.e. coronal equilibrium) and \texttt{CHIANTI} to give us the population fraction of any particular upper state of an ion $\frac{n_u}{n_{\mathrm{ion}}}$ and the population fraction of each ion per element $\frac{n_{\mathrm{ion}}}{n_{\mathrm{element}}}$, an assumed local density $n_e$, and laboratory measurements or theoretical calculations of the atomic data $A_{ul}$ and $\lambda_{ul}$. We follow \citet{DelZanna2002} in using a constant electron pressure $P_e$ to define $n_e (T) = \frac{P_e}{k_B T}$, where $k_B$ is the Boltzmann constant. This single pressure will not be applicable to the entire temperature domain, but errors in the $G(T)$ function can be partially compensated for by the $\Psi(T)$ function as long as the errors are largely a function of temperature and do not vary significantly across lines formed at the same temperature.

The differential emission measure, $\Psi(T)$, describes the density and temperature structure along the line of sight, common to all transitions we observe from the chromosphere, transition region, and corona. Under our assumptions that the ions are predominantly populated by collisions and depopulated by spontaneous emission, the flux observed is proportional to the collision rate. The differential emission measure resembles a reaction rate, $n_e \cdot n_{\textrm{H}}$, weighted by $\frac{ds}{dT}$ which measures how much of the path length $s$ is at a temperature $T$. In emission measure studies of other stars, a volume emission measure is commonly employed that predicts a flux and includes factors of the stellar radius and solid angle filling factor of the emitting plasma. We adopt the line-of-sight approach to be able to compare the DEMs of very different stars to each other and to solar surface features.

For each emission line there is a formation temperature $T_f$ that maximizes the product $G_{{ul}}(T)\cdot \Psi(T)$, and since the emissivity function $G_{ul}(T)$ tends to be very narrowly peaked, the bulk of of the observed line flux is emitted by plasma at $\approx T_f$. By measuring the observed line intensities of transitions with a known $G_{ul}(T)$, we can constrain the value of $\Psi(T)$ within the vicinity of the lines' formation temperatures $T_f$. Amassing a list of observable transitions over a sufficiently wide range of $T_f$ allows us to fit for the parameters of an assumed functional form describing $\Psi(T)$ across the temperature domain of the upper stellar atmosphere. With $\Psi(T)$ in hand and atomic data to construct $G_{ul}(T)$ for the transitions we have not observed but seek to estimate, we can reconstruct the optically thin emission of the chromosphere, transition region, and corona. With the exception of the recombination continua addressed in Section \S\ref{sec:euv_fitting}, optically thin emission lines contribute the majority of the EUV flux from an M dwarf.


\section{Implementation}\label{sec:implementation}
\begin{figure}
    \centering
    \includegraphics[width=\textwidth]{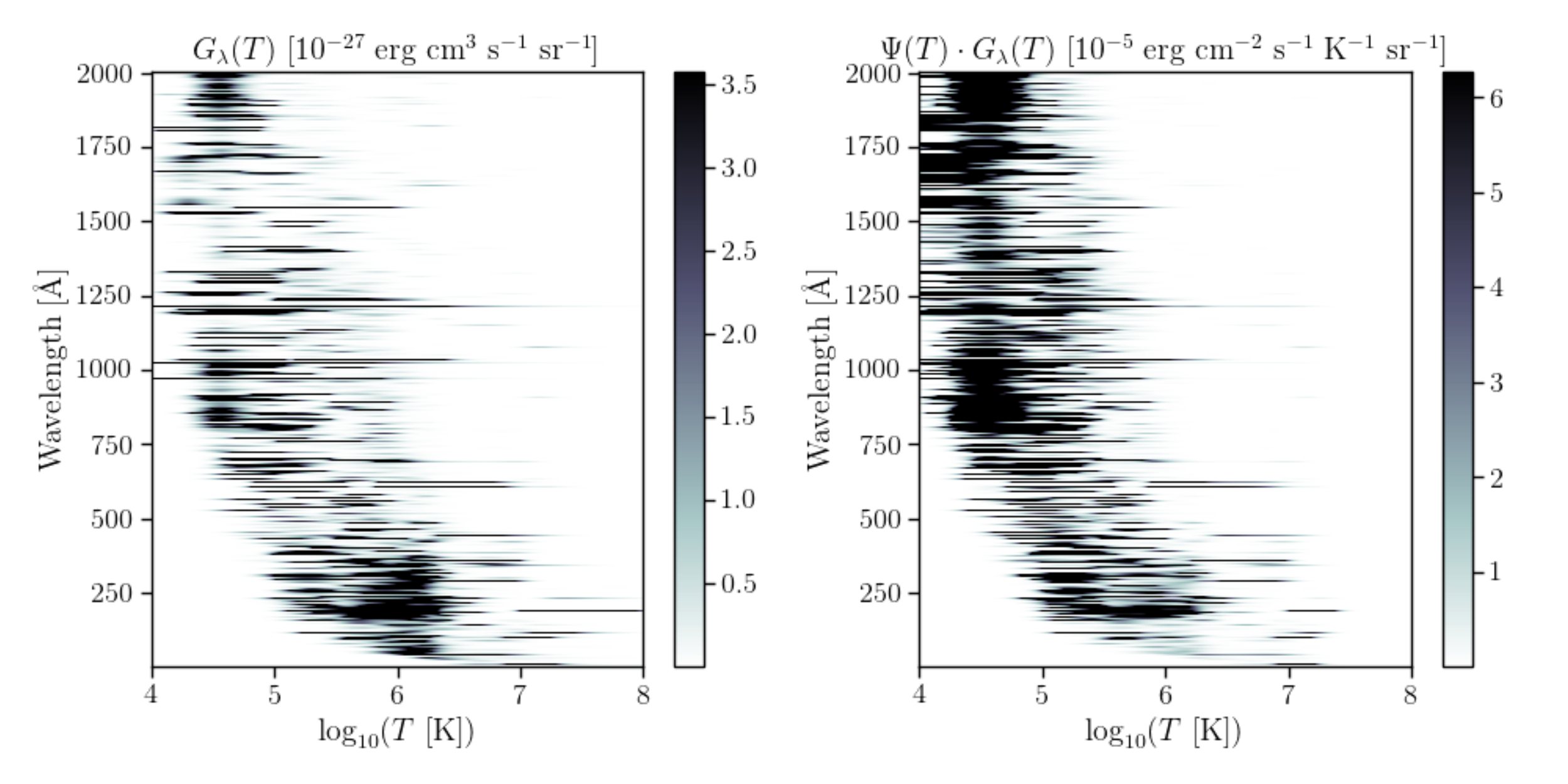}
    \caption{The left panel is a heatmap where the shading represents the value of the emissivity contribution function $G_{\lambda}(T)$ as a function of temperature along the horizontal axis and as a function of wavelength along the vertical axis. This matrix is generated from the \texttt{CHIANTI} atomic database by specifying the atomic abundances and the electron density as a function of temperature $n_{e}(T)$. The right panel multiplies the $G_{\lambda} (T)$ matrix by the DEM, $\Psi(T)$, of the Sun shown as a blue line in Figure \ref{fig:sun_euv_dem}. The colorbars of both panels are cut off at the 60$^{\textrm{th}}$ and 95$^{\textrm{th}}$ percentiles to highlight the strongest lines. While the $G_{\lambda}(T)$ matrix shows the atomic data, the right panel represents the temperature integrand for which lines are actually emitted by stars. EUV lines are largely formed at temperatures between $10^5$ and $10^{6.5}$ K, requiring FUV measurements to constrain the low temperature end and X-ray measurements and/or coronal FUV semi-forbidden transitions of highly ionized iron to constrain the high temperature end.}
    \label{fig:gofnt_heatmap}
\end{figure}

\begin{sidewaysfigure}
    \centering
    \includegraphics[width=1.0\textwidth]{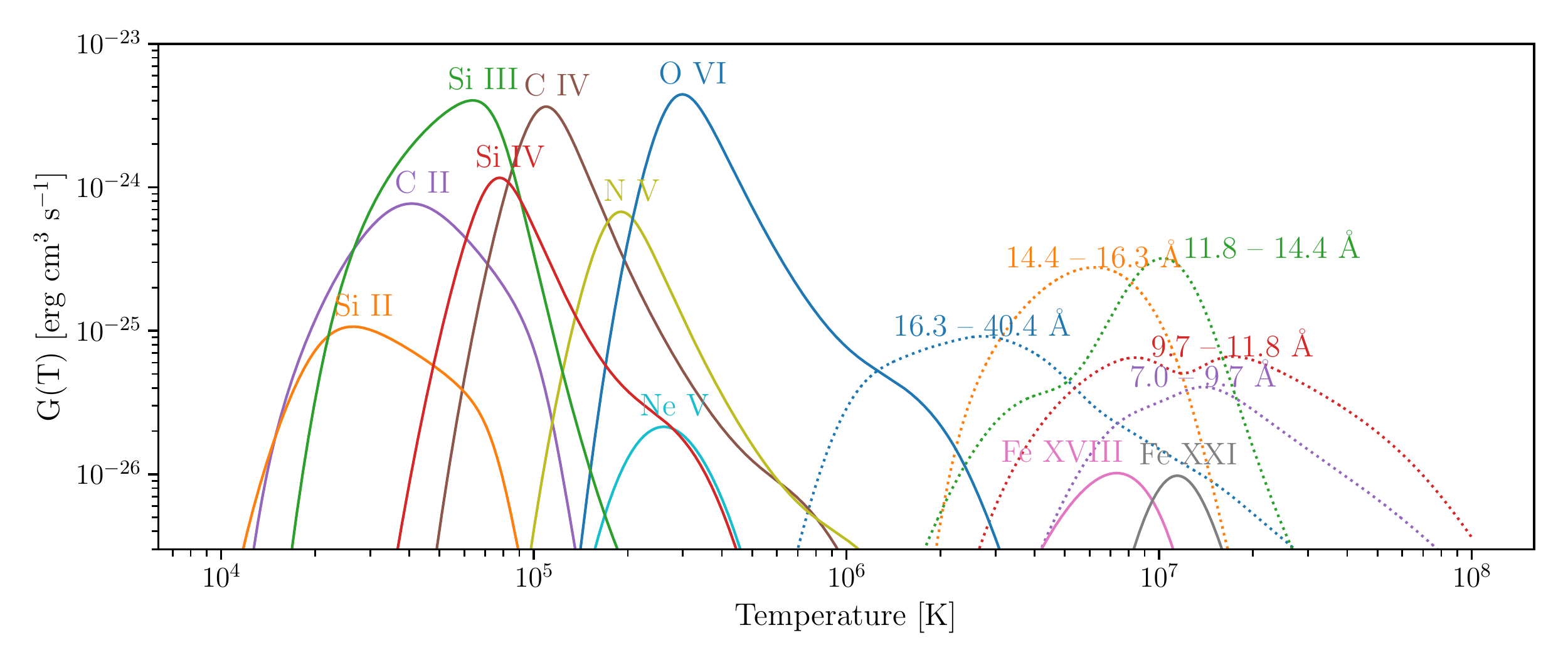}
    \caption{A sample of the emissivity functions $G_{ul}(T)$ and $G_{\lambda}(T)$ described in Sections \S\ref{sec:dem} and \S\ref{sec:implementation}. The solid lines show the summed emissivity functions for some of the strongest transitions of the labeled ion observable by \emph{Hubble} with STIS and COS. The dotted lines show the summed emissivity functions $G_{\lambda} (T)$ for typical wavelength bins in a \emph{Chandra} spectrum of an M dwarf, adding up the $G_{ul}(T)$ functions of all emission lines that fall within the wavelength bin. The ion emissivities are constructed by adding together the emissivities of all lines emitted by that ion, for example both lines of the N V 1239/1243 doublet. The X-ray wavelength bin emissivities are constructed similarly by adding together the emissivities of lines from multiple ions that are emitted at similar wavelengths. This lets us use coarser resolution X-ray spectra for fainter stars but weakens the temperature specificity of individual flux measurements. Note the contribution function for a semi-forbidden transition of \ion{Fe}{21} at 1354 $\textrm{\AA}$ with a formation temperature at $\sim  10^{7}$ K. Its emissivity peaks at a factor of $10^3$ times less than other typically observable FUV lines, making it unobservable for most quiescent M dwarfs. Observing or obtaining an upper limit for the flux of this line provides an additional constraint on $\Psi(T)$ at high temperatures.}
    \label{fig:gofnt_all}
\end{sidewaysfigure}

We use \texttt{CHIANTI 8.0.7} \citep{Dere1997, DelZanna2015} to calculate the $G_{ul}(T)$ functions for all the transitions in the database assuming the elements have a solar coronal abundance \citep{Schmelz2012}. We calculate these emissivity contribution functions across a temperature range from $10^4$ to $10^8$ K for multiple assumed electron pressures ranging from $P_e= 10^{12}$ to $10^{25}$ $k_B$ cm$^{-3}$ K. The majority of coronal emission lines are not strongly sensitive to density, but we test the variation in the predicted flux as a function of the $P_e$ used to calculate $G_{ul} (T)$ in Section \S\ref{sec:pressure_sensitivity}. We assume that $\Psi(T)$ is well-described by

\begin{equation}\label{eq:dem_cheby}
    \log_{10}\Psi(T) = \sum_{n=0}^5 c_n \mathbf{T_n} \left(\frac{\log_{10}T - 6}{2}\right)
\end{equation}
\noindent
where $\mathbf{T_n} (x)$ are the Chebyshev polynomials of the first kind, and their argument is shifted and scaled to transform the domain to the interval $[-1, 1]$. We use the Chebyshev polynomials following the previous work of \citet{Louden2017} and because they form an orthonormal basis. Given the coefficients $c_n$, and the list of emissivity contribution functions $G_{ul} (T)$, we can generate a full spectrum by summing the contribution functions of all emission lines within a wavelength bin centered on a wavelength $\lambda$ with a width $\Delta \lambda$ to get a wavelength-specific contribution function

\begin{equation}\label{eq:wavelength}
    G_\lambda (T) = \sum_{\mathrm{transitions}} G_{ul} (T) \ \mathbf{if} \left[\lvert\lambda_{ul} - \lambda\rvert \leq \Delta \lambda \right].
\end{equation}
\noindent
We then scale the temperature integral of Equation \ref{eq:radiance_dem} to predict the observed flux density in each wavelength bin, assuming the bin is wide enough to contain the entire line profile
\begin{equation}
    F_{\lambda} = \frac{\pi}{\Delta \lambda} \left(\frac{R^2_\star}{d^2}\right) \int_T G_{\lambda} (T) \cdot \Psi(T) \, dT  \; [\mathrm{erg} \, \mathrm{s}^{-1} \, \mathrm{cm}^{-2} \, \text{\AA}].
\end{equation}

The scaling factor assumes that the solid angle emitting the flux is $\approx \pi \left(\frac{R_\star}{d}\right)^2 $ steradians, which is approximate because the corona extends beyond the stellar radius. We create a matrix of $G_\lambda (T)$ with the wavelength axis at a constant resolving power $R = \frac{\lambda}{\Delta \lambda} = 500$ between 1 and 2000 $\textrm{\AA}$, and the temperature axis as 2000 logarithmically spaced points between $10^4$ and $10^8$ K (see Figure \ref{fig:gofnt_heatmap}). Fitting for the coefficients $c_n$ and combining the $\Psi(T)$ model with this matrix allows us to generate a high-resolution spectrum, but since the DEM makes no prescription for line shape, the line profiles are all Dirac-$\delta$ functions, which is why we then divide by the wavelength bin width $\Delta \lambda$ to get the observed flux density. Comparing this model to a real stellar spectrum is only reasonable at a low enough resolution such that the entirety of the line profile is contained within each resolution element. The $R= \frac{\Delta \lambda}{\lambda}=500$ $G_\lambda(T)$ matrix can be downsampled to whatever resolution is required to contain the line widths of any spectral data used for comparison.

By combining either the $G_{ul}(T)$ functions or the $G_{\lambda}(T)$ matrix with the polynomial coefficients $c_n$, we have a generative model for a list of integrated line fluxes or a low-resolution spectrum respectively. To get a usefully constrained model, we need data that covers the full temperature domain. Figure \ref{fig:gofnt_all} shows the $G_{\lambda}(T)$ functions for the wavelength bins of a typical \emph{Chandra} spectrum, where each bin peaks at a slightly different temperature but spans $10^{6}$ and $10^{7.5}$ K, and the $G_{ul}(T)$ functions for the strongest optically thin FUV lines accessible in a \emph{Space Telescope Imaging Spectrograph} (STIS) and \emph{Cosmic Origins Spectrograph} (COS) spectrum from \emph{Hubble}. There is significant overlap in the contribution functions of these transitions near $10^5$ K, but they spread out far enough to constrain $\Psi(T)$ between $10^4$ and $10^{5.5}$ K. Each line flux measurement can be used to derive an average value of $\Psi(T)$ near the formation temperature $T_f$ of the transition

\begin{equation}
    \overline{\Psi} (T_f) = \frac{F_{\textrm{line}}}{\pi \left(\frac{R^2_\star}{d^2}\right)  \int_T G_{ul}(T) dT}
\end{equation}
\noindent
and any individual wavelength bin's flux density can provide a similar constraint by substituting $G_{\lambda} (T)$ in for $G_{ul} (T)$ and dividing by the wavelength bin width $\Delta \lambda$. We do not fit to these averages because we can directly compare our predicted fluxes to the data, but the averages are useful for visualizing how an individual flux measurement constrains the DEM.

Using the affine-invariant Markov Chain Monte Carlo sampler implemented in the Python package \texttt{emcee} \citep{ForemanMackey2013}, we fit the coefficients $c_n$ (see Equation \ref{eq:dem_cheby}) using a combination of the available X-ray data and integrated FUV line fluxes. Since the uncertainties on the emissivities are unknown and we have little a priori information on how to characterize the systematic uncertainties associated with this method, we assume that the variance is boosted by a scaled multiple of the predicted flux, making our log-likelihood

\begin{equation}
    \ln{\mathcal{L}} = \sum_i \ln\left(\frac{1}{\sqrt{2\pi \left(\sigma_{y_i}^2 + \left(s\cdot f(x_i)\right)^2\right)}}\right) - \left(\frac{y_i - f(x_i)}{2\sqrt{\sigma_{y_i}^2 + \left(s\cdot f(x_i)\right)^2}}\right)^2
    \label{eq:likelihood}
\end{equation}
\noindent
 where $f(x_i)$ is the model prediction, $y_i$ is the data, $\sigma_{y_i}$ is the Gaussian uncertainty of the data, and $s$ is the free parameter that characterizes these unknown systematic uncertainties (which are assumed to be independent of the data and temperature). Some contributions to $s$ are likely to be errors in stellar parameters like the stellar abundance, deviations of level populations from true collisional equilibrium or variations in the relative abundances along the line of sight, the departure from being perfectly optically thin $\tau = 0$, and the spatial inhomogeneity of the emitting plasma. This form of the likelihood is independent of the two types of data described above, allowing us to mix together combinations of line fluxes and spectra in different wavelength regimes, so long as we ensure that these do not overlap to count the same data twice.

 We incorporate Bayesian priors on individual parameters to modify the likelihood evaluated in Equation \ref{eq:likelihood}. We sample $\log_{10}(s)$ uniformly between $-2$ and $2$. The mean value of the DEM is set by $c_0$, which is sampled uniformly between 20 to 26 and then exponentially cut off beyond those bounds. These boundary values were chosen to limit the DEM to physical expectations for $10^5 \lesssim n_e \lesssim 10 ^{17}$ cm$^{-3}$, and path-length $10^8 \lesssim ds \lesssim 10^{11}$ cm. The remaining coefficients $c_n$ are sampled uniformly within the bounds $\pm 100$, and then we also require that the base-10 logarithm of the final polynomial be positive at $\log_{10} T = 6$ to prevent unphysically small DEMs and that the derivative be negative at the lower bound $T = 10^4$ K to reflect the higher amount of material in the photosphere compared to the chromosphere. These priors extend generously beyond physically realistic DEM shapes, for example they do not require the DEM to go to 0 at high temperatures, allowing for an infinitely extended corona. Data constrain the parameter distributions to factors of a few at most, with the $s-$factor systematic uncertainty typically restricted to the interval $0.1 < s < 1$.

\section{Testing the DEM Method Against the Sun}\label{sec:sun}
To test our implementation of the DEM on solar data, we use the Solar Irradiance Reference Spectra (SIRS) published by \citet{Woods2009}. This is a disk-integrated spectrum of the quiescent Sun assembled from measurements collected during the 2008 minimum of the solar activity cycle at 1 $\textrm{\AA}$ resolution. Referring to a list of the lines used for the DEM fitting in \citet{Warren1998}, and making a point to select the FUV lines most likely to be detected in \emph{Hubble} observations of M dwarfs, listed in Table \ref{table:sun_lines}, we measure their fluxes in this spectrum by subtracting the continuum and integrating line profiles. Then we selected the X-ray data between $5$ and $50$ $\textrm{\AA}$, comparable to the regions observed by the \emph{Chandra}-\emph{X-ray} \emph{Observatory} and \emph{XMM}-\emph{Newton}, and left the spectrum at its original resolution of 1 $\textrm{\AA}$ wavelength bins, $R =\frac{\lambda}{\Delta \lambda} \leq 50$. This combination of line fluxes and an X-ray spectrum is the same type of data we use for M dwarfs discussed later in this work. Table \ref{table:sun_lines} also lists the integrated fluxes of EUV lines measured from the \citet{Woods2009} spectrum, used in the test described in Section \S\ref{sec:euv_fitting}. The SIRS did not provide error bars, but we assigned errors such that we had three versions of the data with S/N $=$ 1, 10, and 100 to test the sensitivity of the fitting to S/N. The true errors vary across the observations from different instruments and wavelength ranges assembled by \citet{Woods2009}, but never exceed 10\% at instrument native resolutions which are much finer than the 1 $\textrm{\AA}$ bins used here.

\begin{deluxetable*}{cccccc}
\tablecaption{Integrated fluxes of optically thin lines measured in the Solar Irradiance Reference Spectrum \citep{Woods2009} compared to the DEM predictions. \label{table:sun_lines}}
\tablehead{
\colhead{Ion} & \colhead{Wavelengths} & \colhead{$\log_{10} T_f$} & \colhead{Observed Flux} & {FUV/X-ray DEM} & {W/o Anomalous Ions DEM}\\
\colhead{}    & \colhead{[$\textrm{\AA}$]}         & \colhead{$\log_{10}(\textrm{[K]})$}               & \colhead{[$10^{-2}$ erg\,s$^{-1}$\,cm$^{-2}$]} & \colhead{[$10^{-2}$ erg\,s$^{-1}$\,cm$^{-2}$]} & \colhead{[$10^{-2}$ erg\,s$^{-1}$\,cm$^{-2}$]}}
\startdata
\ion{C}{2} & 1335.7 & 4.4 & 14.9 & 11.0 & 11.9\\
\ion{C}{3} & 1175.7 & 4.8 & 4.73 & 7.78 & 3.36\\
\ion{C}{4} \tablenotemark{a}& 1548.2, 1550.7 & 5.0 & 12.0 & 7.95 & 2.17\\
\ion{N}{5} \tablenotemark{a}& 1238.8, 1242.8 & 5.3 & 1.56 & 1.56 & 0.419\\
\ion{Ne}{7}\tablenotemark{b} & 465.2 & 5.7 & 1.47 & 3.07 & 1.60\\
\ion{Ne}{8}\tablenotemark{b} & 770.4, 780.3 & 5.8  & 1.80 & 2.83 & 1.92\\
\ion{O}{3}\tablenotemark{b} & 508.2, 525.8, 599.6, 703.9 & 4.9 & 2.71 & 6.63 &  2.09\\
\ion{O}{4}\tablenotemark{b} & 554.5, 787.7, 790.2 & 5.2 & 5.85 & 19.9 & 4.95\\
\ion{O}{5}\tablenotemark{b} & 629.7, 760.4 & 5.4 & 6.47 & 21.4 & 5.77\\
\ion{Si}{3} & 1206.5 & 4.5 & 6.83 & 22.6 & 18.4\\
\ion{Si}{4}\tablenotemark{a} & 1393.8 & 4.9 & 3.72 & 2.24 & 0.827\\
\ion{Si}{12}\tablenotemark{b} & 499.4 & 6.3 & 0.699 & 0.921 & 1.03\\
\enddata
\tablecomments{In cases where multiple transitions are listed for the same ion, the reported flux is the summed flux across all listed transitions.}
\tablenotetext{a}{These FUV transitions were not used to fit the ``Fit with EUV Lines and without Anomalous Ions" model shown in Figure \ref{fig:sun_euv_dem} as a red solid line.}
\tablenotetext{b}{These EUV transitions were used to fit the ```Fit with EUV Lines and without Anomalous Ions" model shown in Figure \ref{fig:sun_euv_dem} as a red solid line.}
\end{deluxetable*}

\subsection{Pressure Sensitivity}\label{sec:pressure_sensitivity}
\begin{figure}
    \centering
    \includegraphics[width=\textwidth]{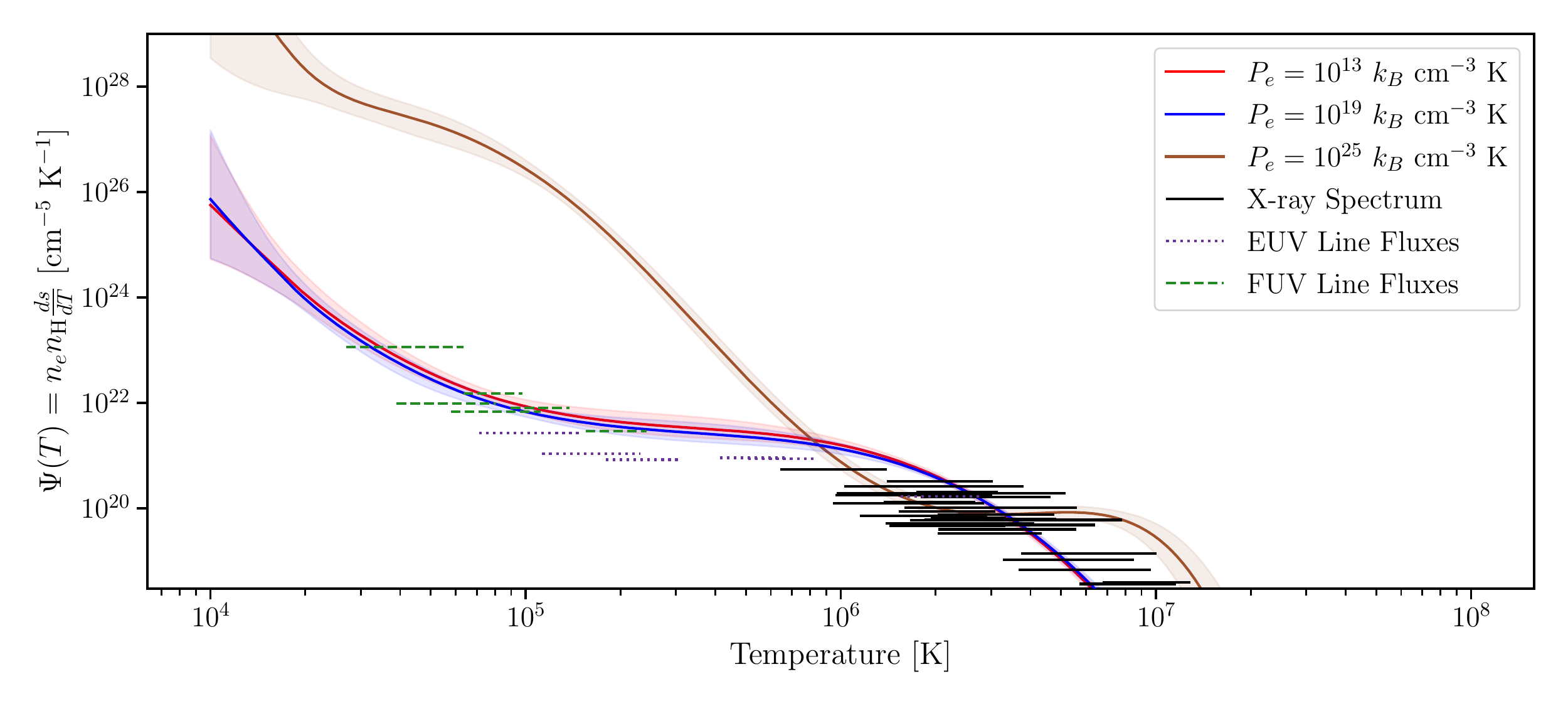}
    \caption{The DEMs of the Sun fit to the same data, a combination of FUV emission lines and X-ray spectra, but with emissivity functions calculated at different electron pressures $P_e$. The horizontal bars represent constraints on the average DEM value $\overline{\Psi}(T_f)$ in the vicinity imposed by the emissivity functions $G(T)$ for ion line fluxes or wavelength bins as described in Section \S\ref{sec:implementation}, with the horizontal extent of the lines representing the full width at half maximum for the emissivity function. These constraints transformed from a flux to an average DEM value are only approximately illustrative of the fit quality. For a true comparison of the DEM prediction to the data used to fit the model, see plots of the predicted line fluxes and X-ray spectra in Figures \ref{fig:sun_euv_lines} and \ref{fig:sun_euv_xray} respectively. The solid lines represent the median DEM value as a function of temperature from the posterior samples for $\Psi(T)$ while the shaded region encloses the 16$^{\textrm{th}}$ to 84$^{\textrm{th}}$ percentile values of the DEM. The red, blue, and brown models correspond to DEM models with emissivities evaluted at electron pressures $P_e = $ $10^{13}$, $10^{19}$, and $10^{25}$ $k_B$ cm$^{-3}$ K respectively. The green dashed bars and black solid bars represent the flux constraints from the FUV lines and X-ray spectra used to fit these models, while the dotted purple bars represent the flux constraints from EUV lines listed in Table \ref{table:sun_lines} but were not used to fit these models. For visualizing these constraints we use the emissivities calculated by assuming the electron pressure $P_e = 10^{19} \, k_B$ cm$^{-3}$.}
    \label{fig:sun_press_dem}
\end{figure}

Across this broad range of temperatures, no single electron density or pressure will accurately describe the environmental conditions of the plasma emitting the observed flux we are using to fit the DEM or the unobserved EUV flux we are trying to predict. However we must assume some function for the electron density $n_e (T)$ to calculate emissivities if we want to fit the DEM at all. Updating the emissivity calculation iteratively would be computationally prohibitive and still fail to accurately describe detailed non-equilibrium physics. By generating multiple emissivity matrices across a broad range of electron pressures, $P_{e} = 10^{12}$ to $10^{25}$ $k_B$ cm$^{-3}$ K, and fitting a DEM to the solar data with each matrix, we test the sensitivity of the DEM shape and calculate the variation in the predicted EUV flux as a function of assumed pressure. Figure \ref{fig:sun_press_dem} shows a representative sample of these DEMs, which vary only slightly for pressures lower than $10^{20}$ $k_B$ cm$^{-3}$ K and are consistent with each other to within 1$\sigma$ variations of the DEM shape. The horizontal lines are the average $\overline{\Psi} (T_f)$ values. To test if any particular model is a statistically significant improvement over the others, we compare the models' values of the Bayesian Information Criterion (BIC, \citealt{Schwarz1978}). \citet{Kass1995} demonstrates that the BIC is related to the natural logarithm of the Bayes factor, such that a $\Delta BIC = 1$ implies the more negative model is $e$ times more likely than the higher one. The BIC is evaluated with the equation
\begin{equation}
    \textrm{BIC} = k \ln(n) - 2 \ln(\hat{\mathcal{L}})
\end{equation}
where $k$ is the number of model parameters, $n$ is the number of datapoints, and $\hat{\mathcal{L}}$ is the maximum-likelihood of the model. This criterion penalizes a higher number of parameters, and the model significance increases as the BIC decreases. All models in this comparison have the same number of parameters, but we also use the BIC later in Section \S\ref{sec:poly_choice} to test our method's sensitivity to polynomial degree.

Table \ref{table:model_significance_pressure} compares each pressure model's BIC, estimated systematic uncertainty characterized by the $s$-factor, and EUV flux integrated from 100 to 912 $\textrm{\AA}$. In the middle of our pressure range, from $10^{17}$ to $10^{20}$ $k_B$ cm$^{-3}$ K, the predicted integrated fluxes are consistent with each other to within 1$\sigma$, but the BIC clearly favors the $10^{19}$ model. We adopt the $P_e = 10^{19}$ $k_B$ cm$^{-3}$ K emissivity matrix for other tests of the Sun DEM model moving forward.  At pressures higher than $10^{21} \ k_B $ cm$^{-3}$ K, the DEM shape and predicted fluxes change drastically, likely because the plasma is optically thick and collisional de-excitation can no longer be ignored. The base of the solar chromosphere is at a pressure of $\sim 10^{20}$ $k_B$ cm$^{-3}$ K \citep{Mariska1992}, so a model DEM that assumes the entire upper atmosphere is at photospheric pressure is bound to be unphysical.

For all other stars, we adopt the same approach of fitting the star's DEM with each pressure separately and choosing the model with the best likelihood. We caution that it is unphysical to interpret these ``best" pressures as representative of a specific region in the stellar atmosphere, and that they should be seen as the most useful average for implementing the DEM and nothing more. Future work could involve testing the DEM with temperature-pressure profiles from stellar atmosphere models to see if this improves the accuracy and precision of the estimated spectrum.

\begin{deluxetable}{ccccc}
\tablecaption{The Bayesian Information Criterion (BIC) for each pressure evaluated against the FUV line fluxes and the X-ray spectrum used to fit the model.\label{table:model_significance_pressure}}
\tablehead{
\colhead{Log Electron Pressure $\log_{10} P_e$ } & \colhead{BIC} & \colhead{Integrated EUV Flux } & \colhead{$\log_{10} s_{\mathcal{L}_{\mathrm{max}}}$} & \colhead{$\log_{10} s$} \\
\colhead{$\log_{10}\left([ k_B \ \mathrm{cm}^{-3} \ \mathrm{K}] \right)$} & \NA & \colhead{[ergs s$^{-1}$ cm$^{-2}$]} & \NA & \NA}
\startdata
$12$ & $-447.8$ & $4.9^{+4.2}_{3.4}$ & $-0.2090$ & $-0.1671^{+0.0739}_{-0.0632}$ \\
$13$ & $-448.6$ & $4.6^{+4.3}_{-3.1}$ & $-0.1852$ & $-0.1699^{+0.0731}_{-0.0626}$ \\
$14$ & $-450.1$ & $4.3^{+4.5}_{-2.8}$ & $-0.1929$ & $-0.1757^{+0.0735}_{-0.0619}$ \\
$15$ & $-449.8$ & $4.5^{+4.2}_{-3.1}$ & $-0.2166$ & $-0.1741^{+0.0726}_{-0.0624}$ \\
$16$ & $-452.3$ & $4.1^{+3.6}_{-2.8}$ & $-0.2249$ & $-0.1783^{+0.0733}_{-0.0631}$ \\
$17$ & $-456.2$ & $3.6^{+3.5}_{-2.3}$ & $-0.2137$ & $-0.1909^{+0.0720}_{-0.0619}$ \\
$18$ & $-460.7$ & $3.5^{+3.3}_{-2.1}$ & $-0.2311$ & $-0.2019^{+0.0697}_{-0.0615}$ \\
$19$ & $-464.7$ & $3.5^{+3.1}_{-2.1}$ & $-0.2424$ & $-0.2178^{+0.0692} _{-0.0596}$ \\
$20$ & $-463.3$ & $3.8^{+3.3}_{-2.4}$ & $-0.2515$ & $-0.2125^{+0.0709}_{-0.0603}$ \\
$21$ & $-455.1$ & $6.2^{+6.1}_{-4.1}$ & $-0.2108$ & $-0.1804^{+0.0729}_{-0.0623}$ \\
$22$ & $-426.2$ & $20^{+20}_{-17}$ & $-0.0881$ & $-0.0591^{+0.0873}_{-0.0725}$ \\
$23$ & $-376.7$ & $59^{+94}_{-59}$ & $0.0828$ & $0.1254^{+0.1199}_{-0.0962}$ \\
$24$ & $-367.6$ & $140^{+220}_{-140}$ & $-0.0113$ & $0.0413^{+0.1069}_{-0.0826}$ \\
$25$ & $-364.1$ & $200^{+370}_{-200}$ & $0.0244$ & $0.0473^{+0.1067}_{-0.0809}$\\
\enddata
\tablecomments{The BIC penalizes model parameters by $k\ln(n)$ where $k$ is the number of parameters being fit and $n$ is the number of data points being fit to. An increasingly negative BIC indicates a better fit. In this case, the most preferred models are the $P_e = 10^{19}$ and $10^{20}$ $k_B$ cm$^{-3}$ K models respectively. We also show the value of $\log_{10} s$ for the maximum likelihood sample from the posterior and the median $\pm 1\sigma$ confidence interval for $\log_{10} s$.}
\end{deluxetable}

\subsection{Sensitivity to S/N}\label{sec:S/N_sensitivity}
\begin{figure}
    \centering
    \includegraphics[width=1.0\textwidth]{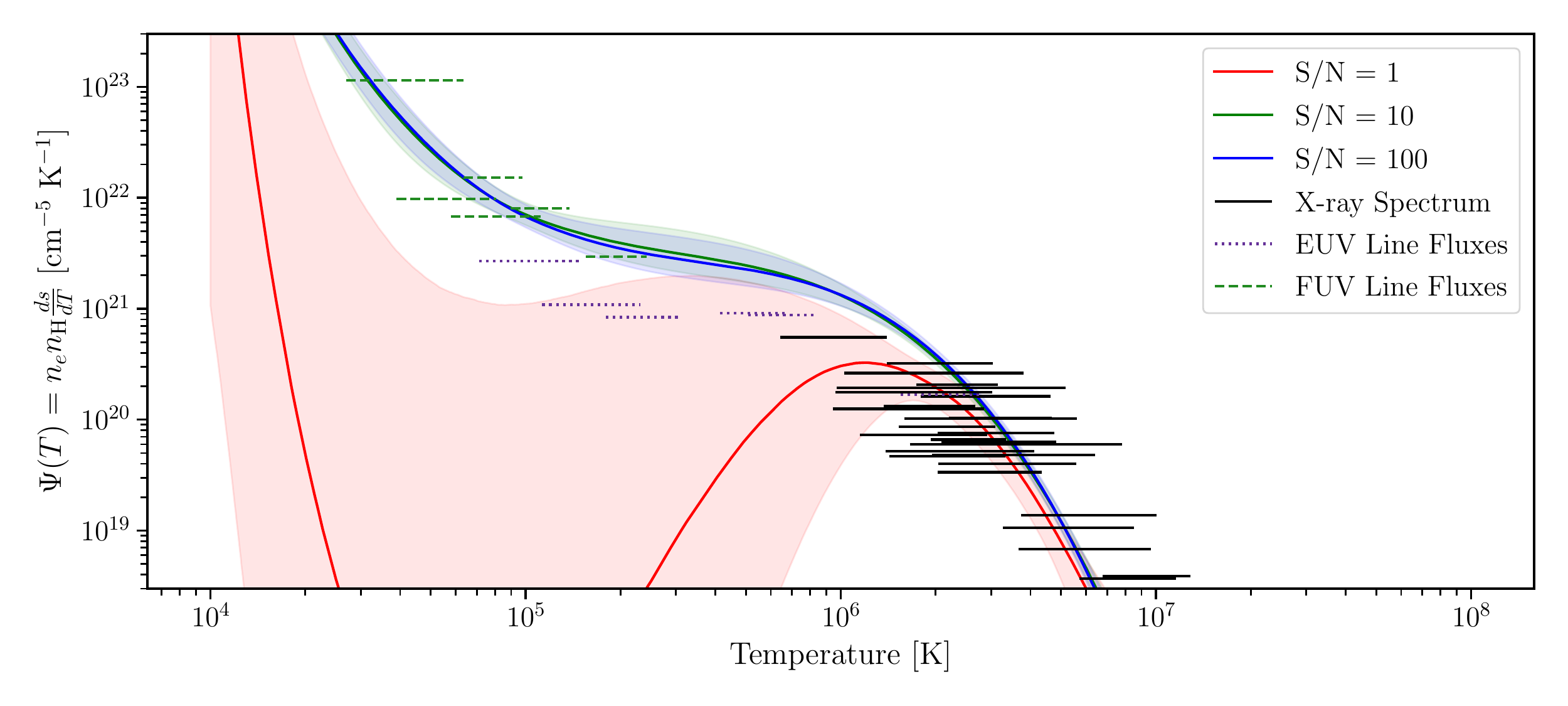}
    \caption{A comparison of fitting the DEM to the same data, a combination of FUV line fluxes and X-ray spectra from the \protect\citet{Woods2009} quiescent Sun spectrum, but with different errors assigned to vary the signal-to-noise ratio (S/N). As shown before in Figure \ref{fig:sun_press_dem}, the solid lines and shaded regions represent the different DEM models while the horizontal bars represent constraints imposed by the flux measurements. The red, green, and blue models represent fitting the DEM at S/N values of 1, 10, and 100 respectively. As shown previously in Figure \ref{fig:sun_press_dem}, the green dashed bars and black solid bars represent the flux constraints from the FUV lines and X-ray spectra used to fit these models. The dotted purple bars represent the flux constraints from EUV lines listed in Table \ref{table:sun_lines} and which were not used to fit these models.} 
    \label{fig:sun_S/N}
\end{figure}
\begin{figure}
    \centering
    \includegraphics[width=\textwidth]{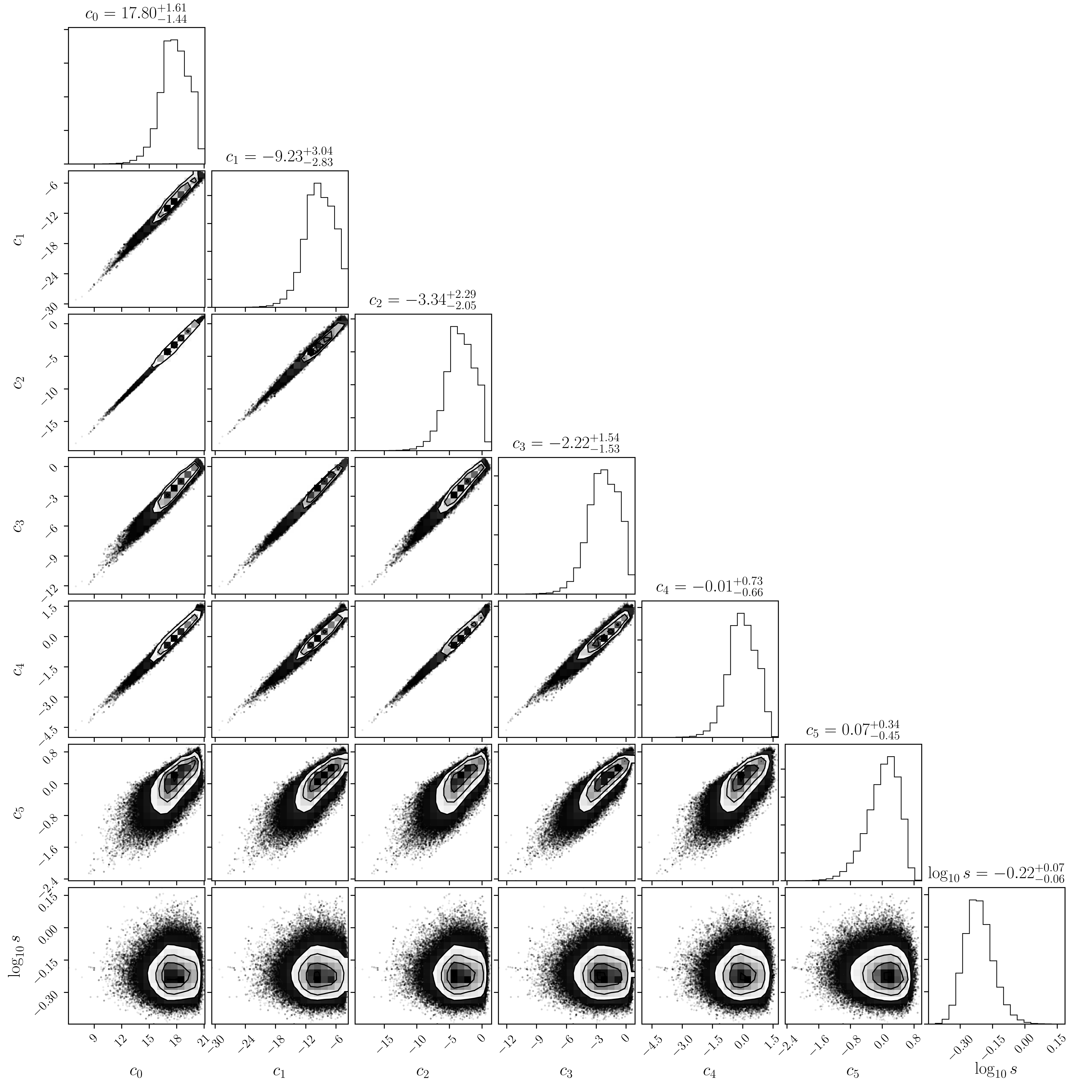}
    \caption{The corner plot for the parameter distributions when fitting the Sun at S/N = 100 shows the median value of $\log_{10} s $ to be $-0.22^{+0.07}_{-0.06}$ which translates to a model uncertainty of $60 \%$ for the predicted flux. This uncertainty dominates the fitting over the uncertainty associated with the data itself.}
    \label{fig:sun_corner}
\end{figure}
Figure \ref{fig:sun_S/N} compares our DEMs fit to the Sun's data with S/N = 1, 10, and 100. For the higher S/N models, the variance is dominated by the uncertainty on the predicted flux, parameterized by the $s-$factor and independent of the shape of the DEM, so changing the S/N of the data used to fit the DEM has little consequence. At S/N = 1, the data uncertainty dominates and the model percentile ranges shift dramatically, with no overlap with the higher S/N models until the higher temperature regimes constrained by many X-ray spectrum fluxes. The low DEM values at FUV temperatures are likely a consequence of the prior requiring a negative derivative at the start of the temperature domain. As mentioned earlier, $s$ is a measure of our average temperature-independent systematic uncertainty that combines the uncertainties on the emissivities with anything else intrinsic to our method's assumptions and approximations. Figure \ref{fig:sun_corner} shows us that $s = 0.63$ in the solar case with S/N$=100$. This indicates that we should assume the systematic uncertainty on any predicted line flux is roughly 60\% of the predicted value. Some of the systematic uncertainty may be attributed to fact that the \citet{Woods2009} spectrum combines observations of quiescence from different instruments taken at different times, which is a problem that will affect most stellar observations and needs to be accounted for in modelling uncertainties.

\subsection{Including EUV Data in the Fitting and Excluding Anomalous Ions}\label{sec:euv_fitting}
With the Sun, we can refer to the EUV lines observed by \citet{Woods2009} in the SIRS, allowing us to see how much information we are losing about the DEM in the stellar case where EUV data is not available. Including the EUV lines gives us more temperature coverage and allows us to exclude three ions from the Na-like and Li-like isolectronic sequences: \ion{N}{5}, \ion{C}{4}, and \ion{Si}{4}. \citet{DelZanna2002} showed that a DEM informed by these ions significantly overpredicts the flux of other transitions because of an anomaly in the \texttt{CHIANTI} ionization equilibrium calculations for these isoelectronic sequences compared to other ions for the same plasma environmental conditions. The factor of discrepancy is not constant across all transitions and densities, so it cannot be corrected for by a consistent known number. This discrepancy constitutes a significant systematic uncertainty that cannot be avoided when fitting the DEM to faint stars with few strong measurable lines that are not from these anomalous ions. Including line fluxes from multiple transitions of other ions can help mitigate the influence of the anomalous ions, but upper limits can still help if the star is too faint to measure these lines.

\begin{figure}
    \centering
    \includegraphics[width=1.0\textwidth]{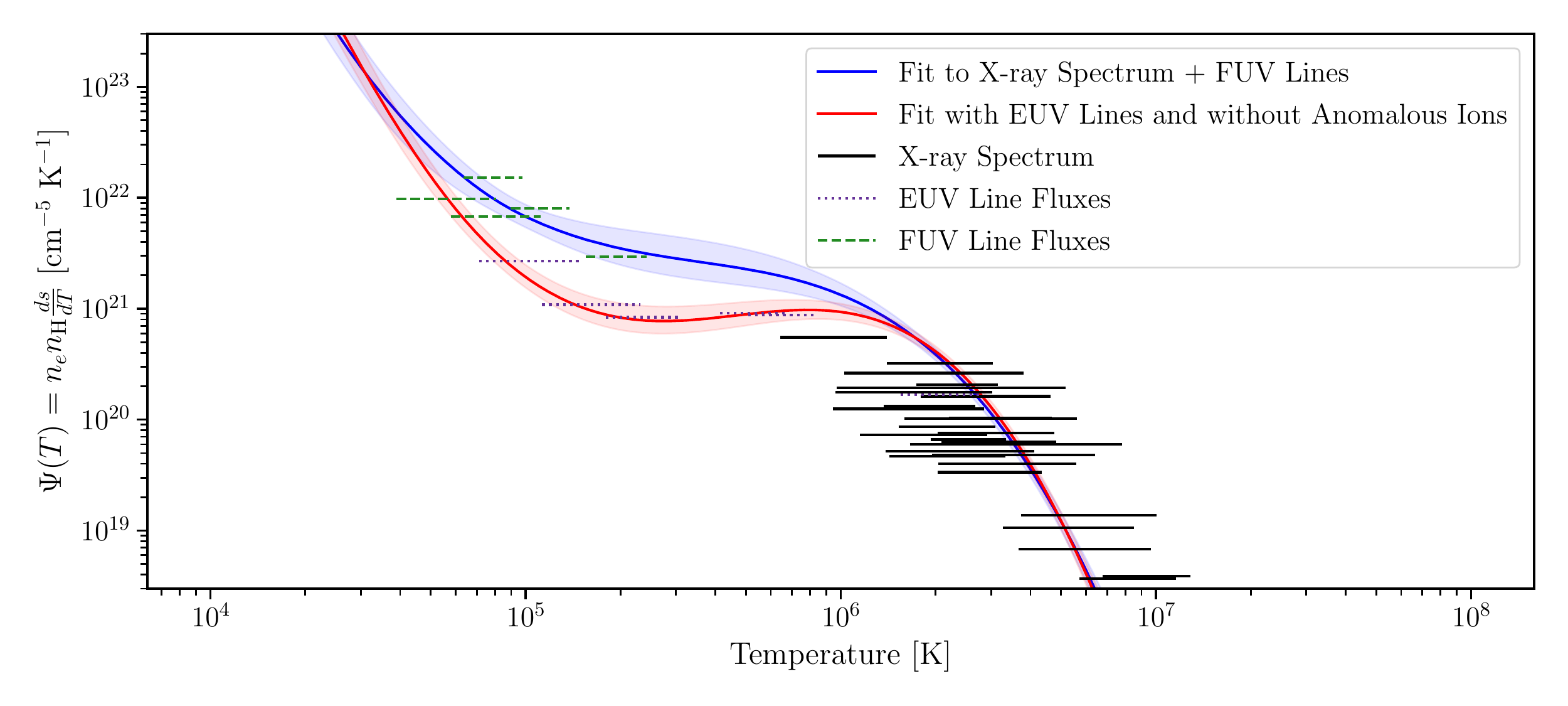}
    \caption{A comparison of the DEM fit using FUV lines and the X-ray spectrum to a new DEM that excludes \ion{N}{5}, \ion{C}{4}, \ion{Si}{4} from the FUV data and includes a number of EUV lines listed in Table \ref{table:sun_lines}. As in Figures \ref{fig:sun_press_dem} and \ref{fig:sun_S/N}, the solid lines and regions represent the median DEM and 1$\sigma$ confidence intervals while the horizontal bars represent constraints imposed by the measured fluxes.}
    \label{fig:sun_euv_dem}
\end{figure}

\begin{sidewaysfigure}
    \centering
    \includegraphics[width=1.0\textwidth]{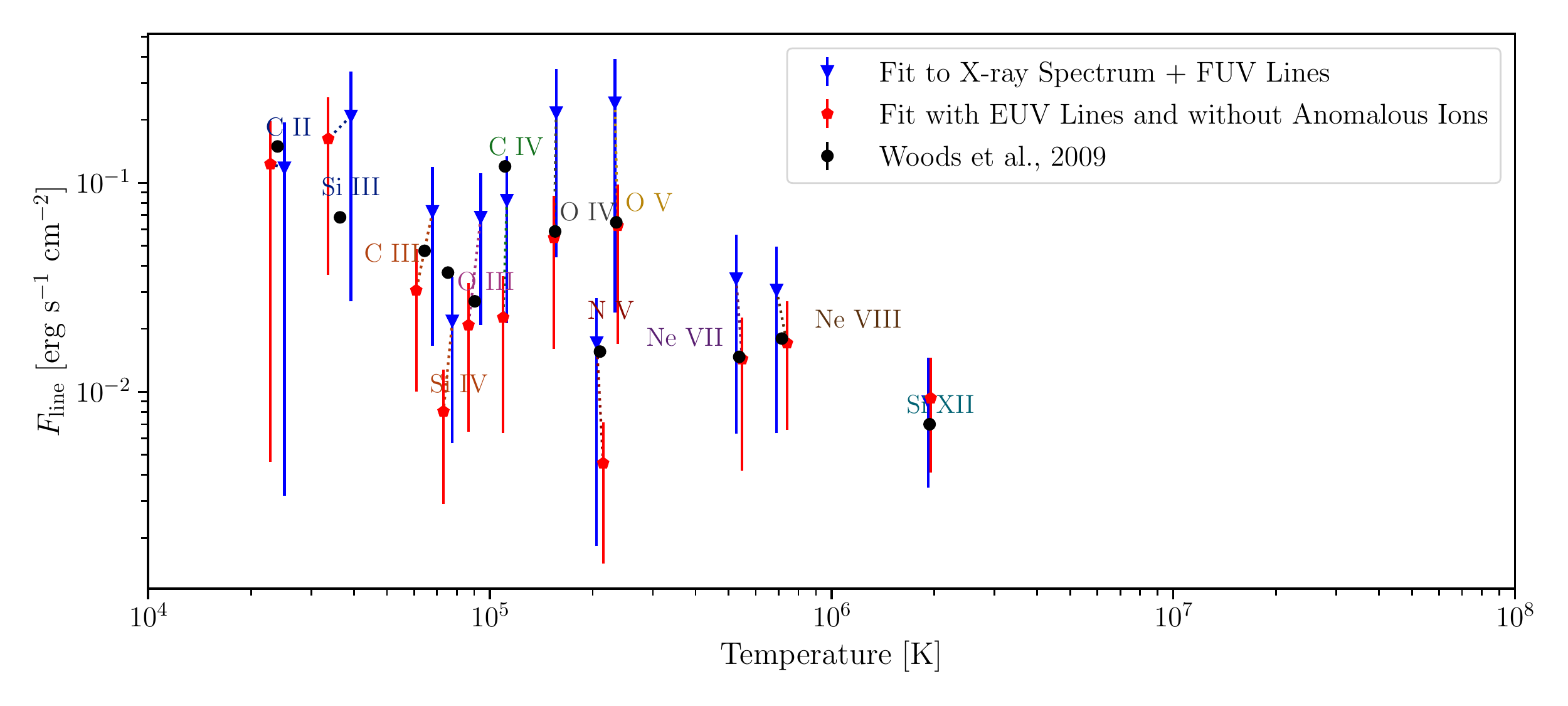}
    \caption{A comparison of the predicted line fluxes from the DEM models shown in Figure \ref{fig:sun_euv_dem} to the lines listed in Table \ref{table:sun_lines}. The color scheme is the same as Figure \ref{fig:sun_euv_dem}, with the red pentagons marking predictions from the model including EUV lines and the blue triangles marking predictions from the model excluding EUV lines. The black points represent the line flux measurements with errorbars for their measurement uncertainties while the model predictions have errorbars for the $s$-factor model intrinsic uncertainty. The high S/N assigned to the \protect\citet{Woods2009} data makes the errorbars nearly invisible. Note how the red model accurately predicts the fluxes of \ion{O}{3}, \ion{O}{4}, \ion{O}{5} formed at roughly the same temperatures as \ion{N}{5}, \ion{C}{4}, \ion{Si}{4} but drastically underestimates the flux of these anomalous ions. Conversely, the blue model is caught in a compromise that slightly underestimates the flux of these ions and overestimates the flux of other FUV ions, but this compromise results in significantly overestimating the flux of EUV ions.}
    \label{fig:sun_euv_lines}
\end{sidewaysfigure}

\begin{figure}
    \centering
    \includegraphics[width=\textwidth]{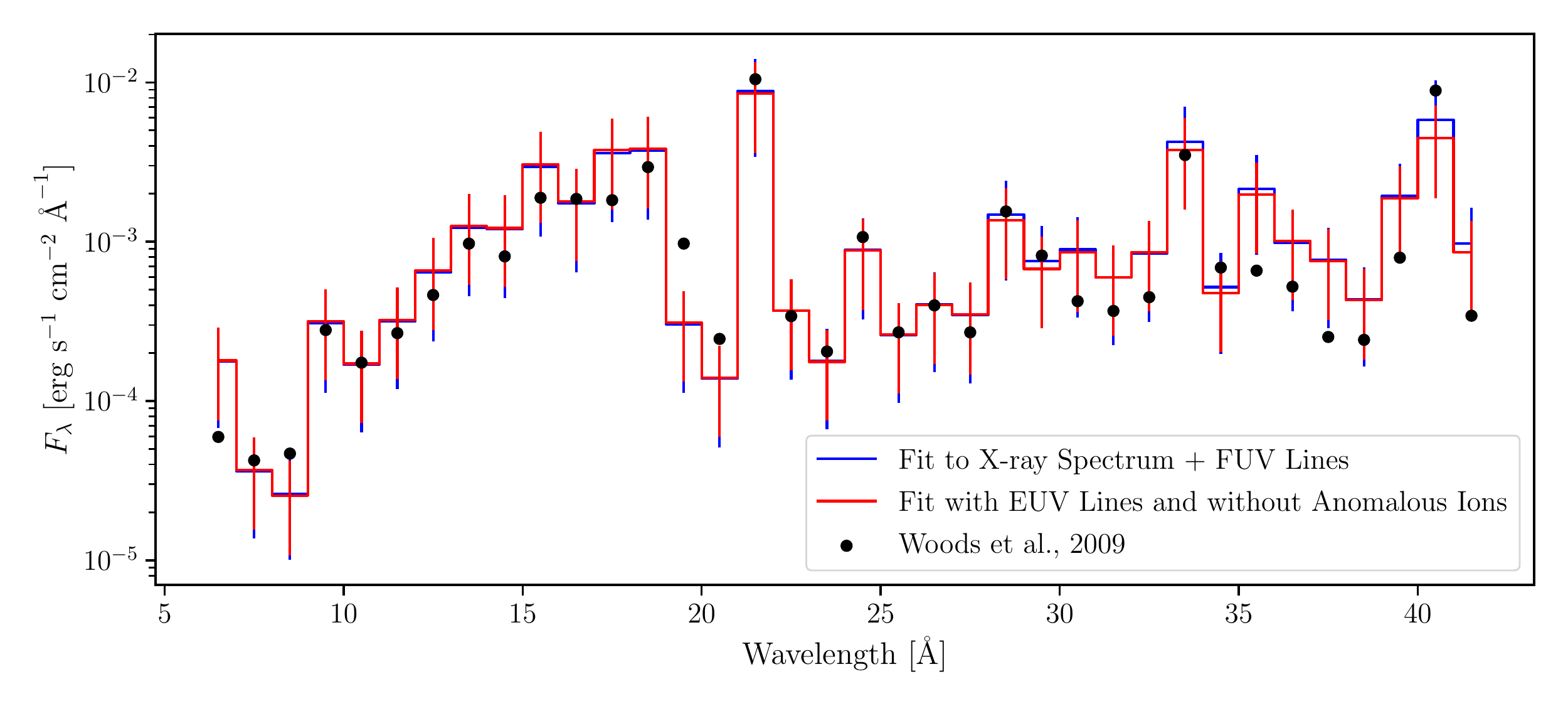}
    \caption{The predicted X-ray spectra from both models discussed in Section \S\ref{sec:euv_fitting} compared to each other and the \protect\citet{Woods2009} SIRS data used to fit the DEM. The solid lines represent models according to the same color scheme as Figure \ref{fig:sun_euv_dem}, blue for the model without EUV lines and red for the model with EUV lines and without the anomalous FUV ions. The errorbars on each model incorporate both the uncertainty in DEM shape and the $s-$factor uncertainty of their respective models in the manner described in Section \S\ref{sec:euv_fitting}. The DEMs do not differ significantly in the temperature regime associated with emitting at X-ray wavelengths, so the spectra for both models are nearly identical. A more complete look at the X-ray spectral data, including data not used to fit the DEM, and our DEM prediction is shown in the top-left panel of Figure \ref{fig:euv_panel}.}
    \label{fig:sun_euv_xray}
\end{figure}

\begin{figure}
    \centering
    \includegraphics[width=\textwidth]{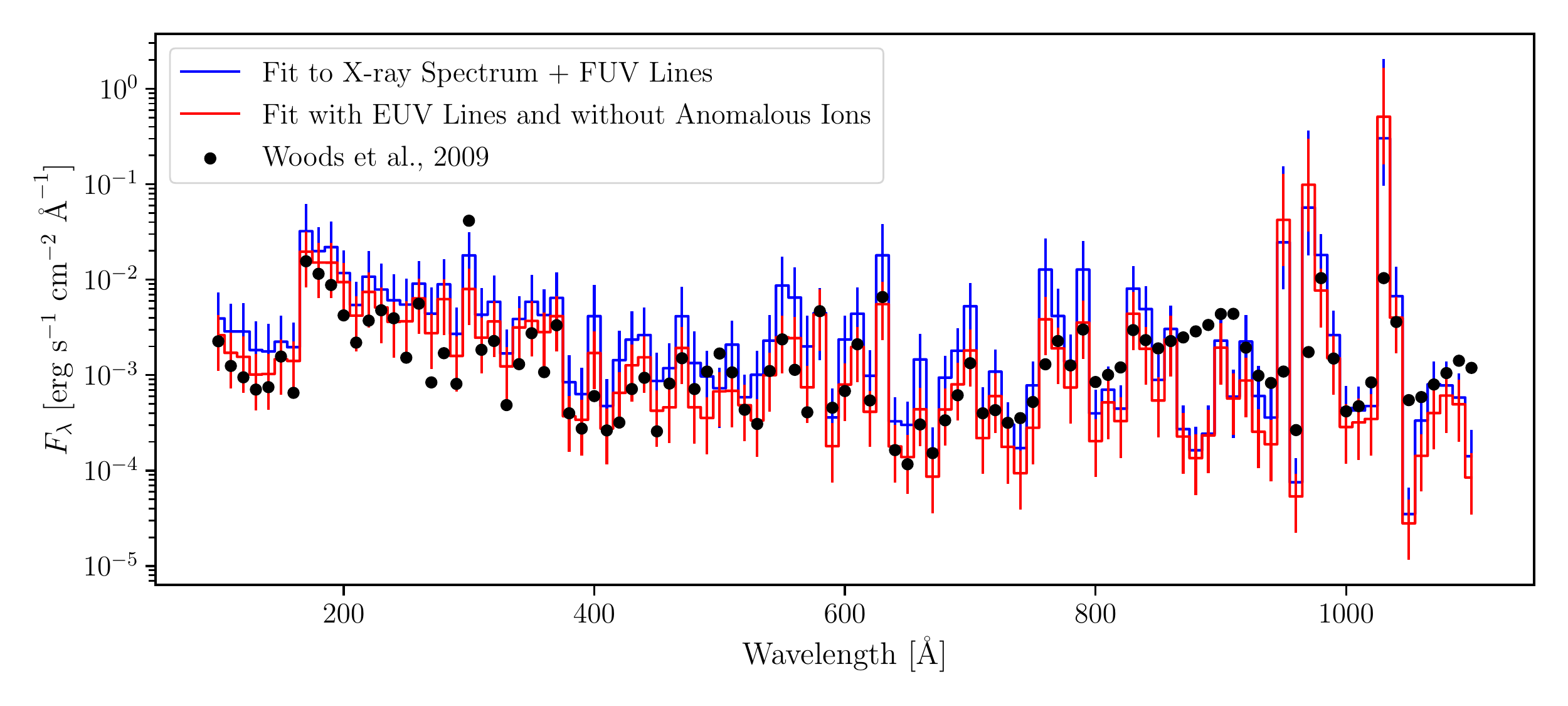}
    \caption{The predicted EUV spectra of both DEM models discussed in Section \S\ref{sec:euv_fitting} with errorbars representing the uncertainties derived in the manner described in that section. The color scheme follows that of Figures \ref{fig:sun_euv_dem} and \ref{fig:sun_euv_xray}, with the solid blue line representing the predicted spectrum and uncertainty of the model fit without EUV lines and the solid red line representing the predicted spectrum and uncertainty of the model fit with EUV lines and without the anomalous ions. The black points are the data from \protect\citet{Woods2009}. Both model spectra show nearly the same spectral shape as the data, excepting the ramp feature at wavelengths between 800 and 912 $\textrm{\AA}$, discussed in Section \S\ref{sec:euv_fitting}. The blue model consistently overestimates the spectrum although the 1$\sigma$ errorbars usually include the data.}
    \label{fig:sun_euv_euv}
\end{figure}

To help characterize the magnitude of these discrepancies, we include EUV lines from 7 ions that are currently unobservable for our M dwarf sample of interest, listed in Table \ref{table:sun_lines}, and drop the anomalous ions (retaining some of the FUV lines and the X-ray spectrum) when fitting the model labeled ``Fit with EUV Lines and without Anomalous Ions" in Figures \ref{fig:sun_euv_dem}, \ref{fig:sun_euv_lines}, \ref{fig:sun_euv_xray}, and \ref{fig:sun_euv_euv}. This new DEM model shifts down by a factor of $\sim 5$ in between $10^5$ and $10^6$ K (see Figure \ref{fig:sun_euv_dem}) to match the EUV line fluxes (see Figure \ref{fig:sun_euv_lines}) that are not informing our stellar-analogous DEM. The DEMs agree with each other at the higher temperatures constrained by the X-ray spectra resulting in nearly identical predictions in that spectral regime (see Figures \ref{fig:sun_euv_dem} and \ref{fig:sun_euv_xray}). The predicted line fluxes from both models are compared to the data in Table \ref{table:sun_lines} and Figure \ref{fig:sun_euv_lines}, and highlight the problem of the anomalous ions. The model without EUV lines and including the anomalous ions predicts the FUV fluxes reasonably well, adopting a compromise position between FUV lines formed at similar temperatures that have discrepant DEM constraints (see Figure \ref{fig:sun_euv_dem}), but this compromise still overestimates the flux of the EUV lines by up to a factor of 5. When applying the DEM to M dwarfs without EUV data, we include the ions with anomalous \texttt{CHIANTI} emissivities because these are the strongest lines available and we cannot afford to simply exclude them. Measuring upper limits for the fluxes of other transitions formed at similar temperatures can mitigate the influence of the anomalous ions when combined with the $s-$factor uncertainty, as demonstrated in our modeling of AU Mic in section \ref{sec:au_mic}.

In Figure \ref{fig:sun_euv_lines}, the error bars associated with the plotted data point incorporate both the posterior distributions of the DEM shape and the $s-$factor uncertainty of the ``Fit with EUV Lines and without Anomalous Ions" model. We do this by drawing randomly from the posterior sample of the MCMC fitting to generate a sample $\Psi(T)$ using the Chebyshev coefficients $c_n$ (see Equation \ref{eq:dem_cheby}) with an associated $s-$factor. The $\Psi(T)$ is combined with $G(T)$ to predict the flux of an observed data point, $y_i$, giving a model flux $f(x_i)$. Multiple sample draws in this fashion would only represent the uncertainty associated with the DEM shape and exclude the $s-$factor. To include the model-intrinsic uncertainty, we draw randomly from the Gaussian distribution $\mathcal{N}(\mu=f(x_i), \sigma={s\cdot f(x_i)})$ and record the prediction from the flux distribution parameterized by a single MCMC posterior sample. This process is executed with $N = 5 \times 10^4$ draws from the model parameters' posterior distribution to describe the full range of the model's predicted flux.  The errorbar for a particular datapoint represents the width of the 16$^{\textrm{th}}$ to 84$^{\textrm{th}}$ percentile interval for this distribution built up of random draws. Figures \ref{fig:sun_euv_xray} and \ref{fig:sun_euv_euv} show the uncertainties of both models as errorbars on the models' respective predicted spectra using the same method. The $s-$factor dominating the uncertainty results in errorbars that scale according to the magnitude of the flux predicted by the model.

When we compare the predicted EUV flux of the Sun from both fits to the observed spectrum itself, the DEM prediction fit without EUV lines overestimates the data by 80\% (see Figure \ref{fig:sun_euv_euv} and Table \ref{table:all_euv_fluxes}). The $s-$factor uncertainty for this model estimates that each line's predicted flux has an uncertainty $60\%$ of the predicted value, so the 1$\sigma$ confidence interval of the model still encompasses the observed data. The DEM prediction including EUV lines and excluding anomalous ions underestimates the integrated flux by only $0.01\%$, but does have a significant $57\%$ $s-$factor uncertainty. Within the EUV regime, there are 3 different recombination continuum regions that form from excess kinetic energy emitted when an ion captures a free electron into a bound state. Only one of them, the \ion{H}{1} 912 $\textrm{\AA}$ continuum, is a significant contributor to the total EUV flux integrated from 100 to 912 $\textrm{\AA}$, accounting for 15\% in the \citet{Woods2009} spectrum of the Sun. The other two regions, \ion{He}{2} 229 $\textrm{\AA}$ and \ion{He}{1} 504 $\textrm{\AA}$, contribute 3\% and 2 \% respectively. This falls within our uncertainties on the predicted flux, but it would be worth investigating if it is possible to incorporate these recombination continua in the DEM model without added parameters. The reconstruction of the EUV spectrum only used the emissivities of optically thin emission lines in the $G_{\lambda} (T)$ emissivity matrices and does not account for any contribution from continuum processes. The data required to incorporate free-bound, free-free, and two-photon continua exist in \texttt{CHIANTI}, but as of this work we have not included these sources of emissivity in the $G_{\lambda} (T)$ emissivity matrices. In future work we hope to include these processes for both fitting the DEM to X-ray spectra and reconstructing the EUV spectrum.

\subsection{Abundance Sensitivity}\label{sec:abundance_sensitivity}
We use the solar coronal abundances from \citet{Schmelz2012} stored in \texttt{CHIANTI} and an ionization equilibrium calculated at those abundances to get the $G_{ul}(T)$ functions for each emission line. Then for every $G_{ul}(T)$ function, we multiply by the stellar abundance if the ion's atomic number $Z > 2$. A higher metal abundance should shift the ionization equilibrium, which would have some effect on the self-consistency of the emissivity calculations. To verify whether or not this effect would be significant, we tested how the DEM fitting was sensitive to abundance in this crude framework. When we fit to stars other than the Sun we can refer to the literature and use the best abundance available, but we should not expect to always have an accurate and precise abundance measurement for the stars we are fitting the DEM to. Fitting the Sun with emissivity matrices generated using a super-solar [Fe/H] $= 1$ and a sub-solar [Fe/H] $= -1$ abundance resulted in the overall DEM shifting up or down to compensate. The predicted EUV fluxes obtained by combining each model DEM with their respective emissivity matrix differed by $< 15\%$.

This makes sense since the $G_{ul}(T)$ of an emission line from any metal is linearly proportional to the abundance. In fact, this harks back to the original implementation of the emission measure distribution in \citet{Pottasch1963}, where the author determined the relative abundances of elements by shifting them around to minimize the scatter in the emission measure distribution. This abundance adjustment is contingent on the assumption of the relative abundances being constant throughout the optically thin plasma, but everything else about this implementation of the DEM hinges on this assumption as well. This will also let us compare the DEM shapes of stars with different abundances by modifying the temperature independent coefficient $c_0$ to shift the overall DEM shape, marginalizing over the stellar abundance for any future study of the DEMs of a population of stars. However, if the relative abundances of an M dwarf corona differ from the Sun, the ions of an affected element will be consistently discrepant from the final DEM fit, and this is likely our greatest source of systematic uncertainty for stars other than the Sun. In the Sun, heavy elements with a first ionization potential $< 9$ keV tend to be enhanced in the corona relative to the photosphere \citep{Drake1995}. There is evidence to suggest that this first ionization potential effect varies with spectral type and may even be reversed in M dwarfs \citep{Drake1995, Wood2012,Laming2015}. 

A single line from a species constrains the DEM near its ionization temperature, but many lines from an element across multiple ionization stages can constrain the element's relative abundance by fitting an element-specific DEM and shifting it to match the DEM fit without that element. This will only be possible for the brightest stars and the C, O, or Si species, since this requires bright lines from at least three ionization stages formed at conditions valid for the DEM method. Fitting DEMs for as many nearby main-sequence stars as possible may allow us to generate this stellar coronal abundance library. A future and intermediate improvement would be to use the sample of stars for which \emph{EUVE} allowed papers like \citet{Drake1995} to determine coronal abundances and select the most appropriate analog for a target we are fitting the DEM to. We intend to use this approach for future work.

\subsection{Polynomial Degree}\label{sec:poly_choice}
\begin{figure}
    \centering
    \includegraphics[width=1.0\textwidth]{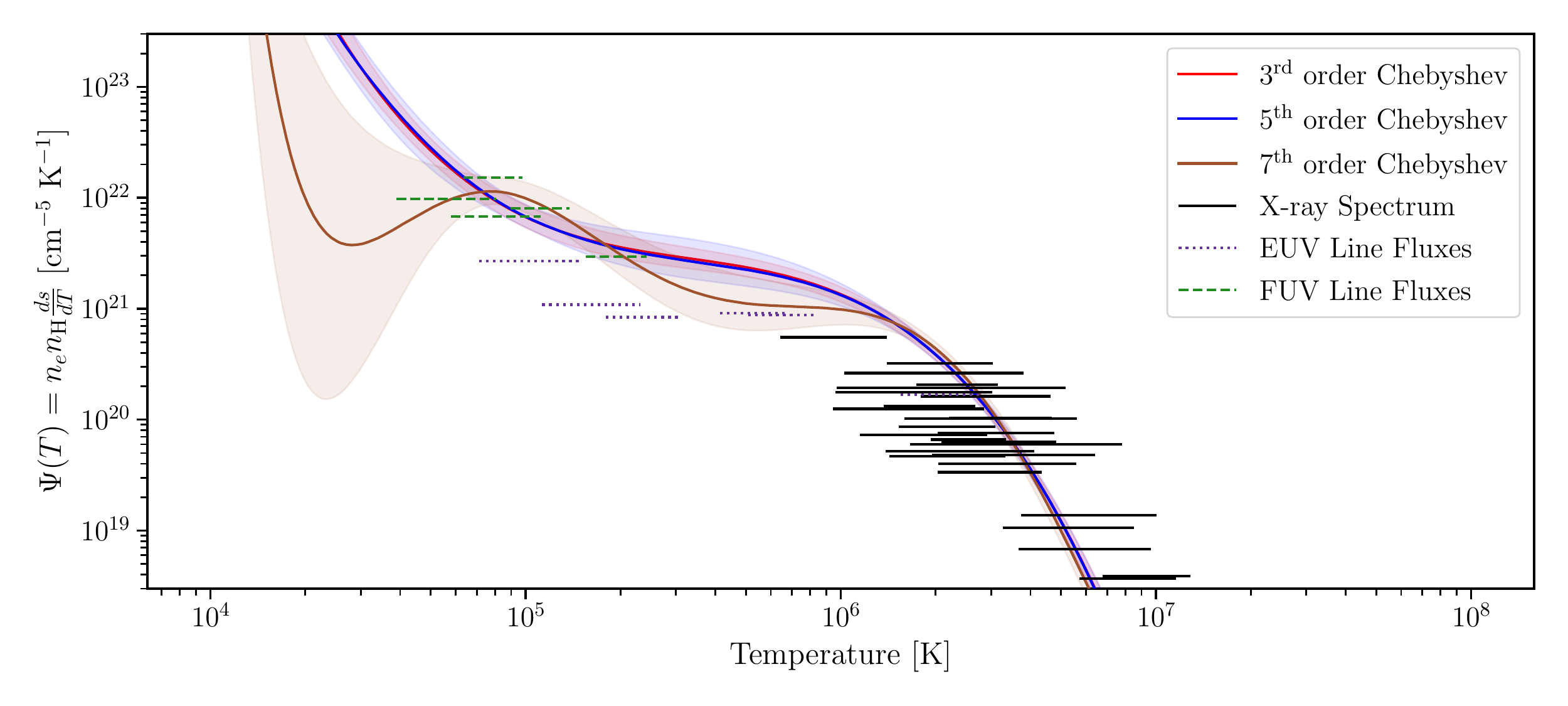}
    \caption{This plot compares the Sun's DEMs fit with different polynomial orders to each other and the flux constraints imposed by the observed FUV lines and X-ray spectrum used to fit the data according to the same scheme described in the caption of Figure \ref{fig:sun_press_dem}. The red model is a 3$^{\rm{rd}}$ order polynomial, the blue model is the 5$^{\textrm{th}}$ order polynomial we adopt as our standard approach, and the green is a 7$^{\rm{th}}$ order polynomial. The spread in FUV line formation temperatures sets a slope for the DEM to match between $2 \times 10^4$ and $2 \times 10^5$ K, while the X-ray spectrum sets the slope between $10^6$ and $2 \times 10^7$ K. Table \ref{table:model_significance_polynomial} compares the different order models to show that the different models predict consistent EUV fluxes.}
    \label{fig:sun_poly_dem}
\end{figure}

Higher polynomial degrees allow more flexibility between the well-constrained temperatures. The spread of FUV and X-ray formation temperatures constrains the slopes in their respective regimes ($2 \times 10^4$ K $<$ FUV $< 2 \times 10^5$ K, $10^6$ K $<$ X-ray $< 2 \times 10^7$ K), and this rigidly constrains lower order polynomials. The models show the most agreement with each other around $T = 10^5$ K, where there are a number of FUV lines with overlapping emissivities to anchor the fit (see Figures \ref{fig:gofnt_all} and \ref{fig:sun_poly_dem}), while there are significant discrepancies at temperatures lower than the constraints imposed by the FUV lines and higher than the constraints imposed by the X-ray spectrum. Table \ref{table:model_significance_polynomial} compares the BIC values for the different polynomial order and since the Sun's DEM shows very little curvature, the BIC prefers the 3$^\textrm{rd}$ order model and penalizes the complexity of higher order polynomials. Table \ref{table:model_significance_polynomial} also shows that the models predict values consistent within 1 $\sigma$ for the EUV flux integrated between 100 to 912 $\textrm{\AA}$, and the $s-$factor uncertainty decreases slightly for higher order polynomials. This consistency is a product of both the low curvature of the Sun's DEM and the fact that the FUV and X-ray data constrain either end of the temperature interval responsible for EUV emission lines. When we ran a similar test with AU Mic, we found that orders below 5 were unable to match the complexity of the data, so we move forward with the 5$^\textrm{th}$ order as our standard approach for fitting other stars. Most targets will have too few data points to merit a model with many more parameters than a 5$^{\textrm{th}}$ order polynomial.

\begin{deluxetable}{ccccc}
\tablecaption{The Bayesian Information Criterion (BIC) for each polynomial model.\label{table:model_significance_polynomial}}
\tablehead{
\colhead{Chebyshev Polynomial Order} & \colhead{BIC} & \colhead{Integrated EUV Flux } & \colhead{$\log_{10} s_{\mathcal{L}_{\mathrm{max}}}$} & \colhead{$\log_{10} s$}\\
\NA & \NA & \colhead{[ergs s$^{-1}$ cm$^{-2}$]} & \NA & \NA}
\startdata
$3$ & $-472.0$ & $3.8^{+2.6}_{-2.3}$ & $-0.2438$ & $-0.2182^{+0.0696}_{-0.0600}$ \\
$4$ & $-468.4$ & $3.7^{+2.7}_{-2.3}$ & $-0.2516$ & $-0.2173^{+0.0690}_{-0.0610}$ \\
$5$ & $-464.7$ & $3.5^{+3.1}_{-2.1}$ & $-0.2515$ & $-0.2177^{+0.0696}_{-0.0603}$ \\
$6$ & $-464.7$ & $3.4^{+2.9}_{-2.1}$ & $-0.2372$ & $-0.2208^{+0.0692}_{-0.0597}$ \\
$7$ & $-464.9$ & $3.0^{+2.4}_{-1.8}$ & $-0.2662$ & $-0.2348^{+0.0706}_{-0.0599}$ \\
$8$ & $-461.1$ & $2.9^{+2.4}_{-1.8}$ & $-0.2852$ & $-0.2350^{+0.0698}_{-0.0600}$ \\
$9$ & $-457.1$ & $2.9^{+2.5}_{-1.8}$ & $-0.2709$ & $-0.2337^{+0.0709}_{-0.0593}$ \\
$10$ & $-454.0$ & $3.1^{+3.0}_{-1.9}$ & $-0.2858$ & $-0.2307^{+0.0717}_{-0.0616}$ \\
\enddata
\tablecomments{The Sun's DEM shows very little complexity, so the 3$^{\textrm{rd}}$ order polynomial model is most preferred, with each subsequent order scoring a worse BIC. This was not true for the AU Mic test for which orders below 5 were too inflexible to match the constraints of the prior and the data.}
\end{deluxetable}

\section{Applying Our DEM Method to AU Mic}\label{sec:au_mic}
AU Mic is a nearby young and active M dwarf with an observable debris disk, $9.979 \pm 0.04$ pc away and $22 \pm 3$ Myr old \citep{MacGregor2013, Bustos2019, Plavchan2020}. The system also hosts at least one confirmed planet and possibly a second planet candidate \citep{Plavchan2020}, making it a rare case of an M dwarf planetary system with a precisely known young age. Despite being nearby, and having been observed multiple times with \emph{EUVE}, the quiescent EUV spectrum of AU Mic is poorly constrained (see Section \ref{sec:final_model_spectra} and Figure \ref{fig:euv_panel}), so some method of reconstructing the EUV spectrum is required to study this planetary system in detail. On the other hand, the X-ray and FUV data for this star are extremely precise considering how intrinsically faint the star is, allowing us to fit a very well-constrained DEM and compare our implementation to earlier work published in \citet{DelZanna2002}. \citet{Pagano2000} published a very thorough list of emission line fluxes measured from spectra taken during quiescence, a subset of which we use to fit the DEM for AU Mic and list in Table \ref{table:fuv_lines}. \citet{Redfield2002} published separate quiescent and flare FUSE measurements of FUV lines and \citet{Redfield2003} reported the quiescent coronal line fluxes listed in Table \ref{table:fuv_lines}. The bandpass for \emph{FUSE} overlaps with that of COS, so we fit separate DEMs including and excluding the \emph{FUSE} measurements to demonstrate the usefulness of transitions observed between $900$ and $1100$ $\textrm{\AA}$ in constraining the high-temperature end of the DEM, motivating future COS observations of cool dwarfs.

The majority of the lines listed in Table \ref{table:fuv_lines} will be unobservable for other fainter cool dwarfs, but having the ground truth of which lines are emitted from the upper atmospheres of cool dwarfs is immensely useful for future DEM fitting. If a line is observed for AU Mic, we can place upper limits on the flux from that line for another star, constraining the DEM near that line's formation temperature. We fit the DEM assuming solar coronal abundances from \citet{Schmelz2012}, using the FUV lines listed in Table \ref{table:fuv_lines} and an  X-ray spectrum from the Reflection Grating Spectrometer (RGS, \citealt{denHerder:2001})  on \emph{XMM-Newton}, observed in October 2018 \citep{Kowalski2019AAS}. The spectrum was resampled at 1 $\textrm{\AA}$ resolution before fitting. The lightcurve for this observation showed multiple flares and we use the quiescent X-ray spectrum from the work of Kowalski et al. (in prep) and Tristan et al. (in prep). This observation was not concurrent with the data obtained by \citet{Pagano2000} and \citet{Redfield2003}, so it is possible that the X-ray and FUV data are not from identical levels of quiescence. Very few targets will be likely to have concurrent X-ray and FUV observations, so this issue will plague most of the stars needing EUV reconstruction. We compare our fit using FUV lines and an X-ray spectrum to another fit using the same FUV lines and X-ray line flux measurements reported by \citet{Wood:2018}. We also note that whenever possible, a line flux measurement is more useful than a spectral bin, and if high temperature emission line strengths can be measured, these should be favored over the use of spectral bins.
\begin{figure}
    \centering
    \includegraphics[width=1.0\textwidth]{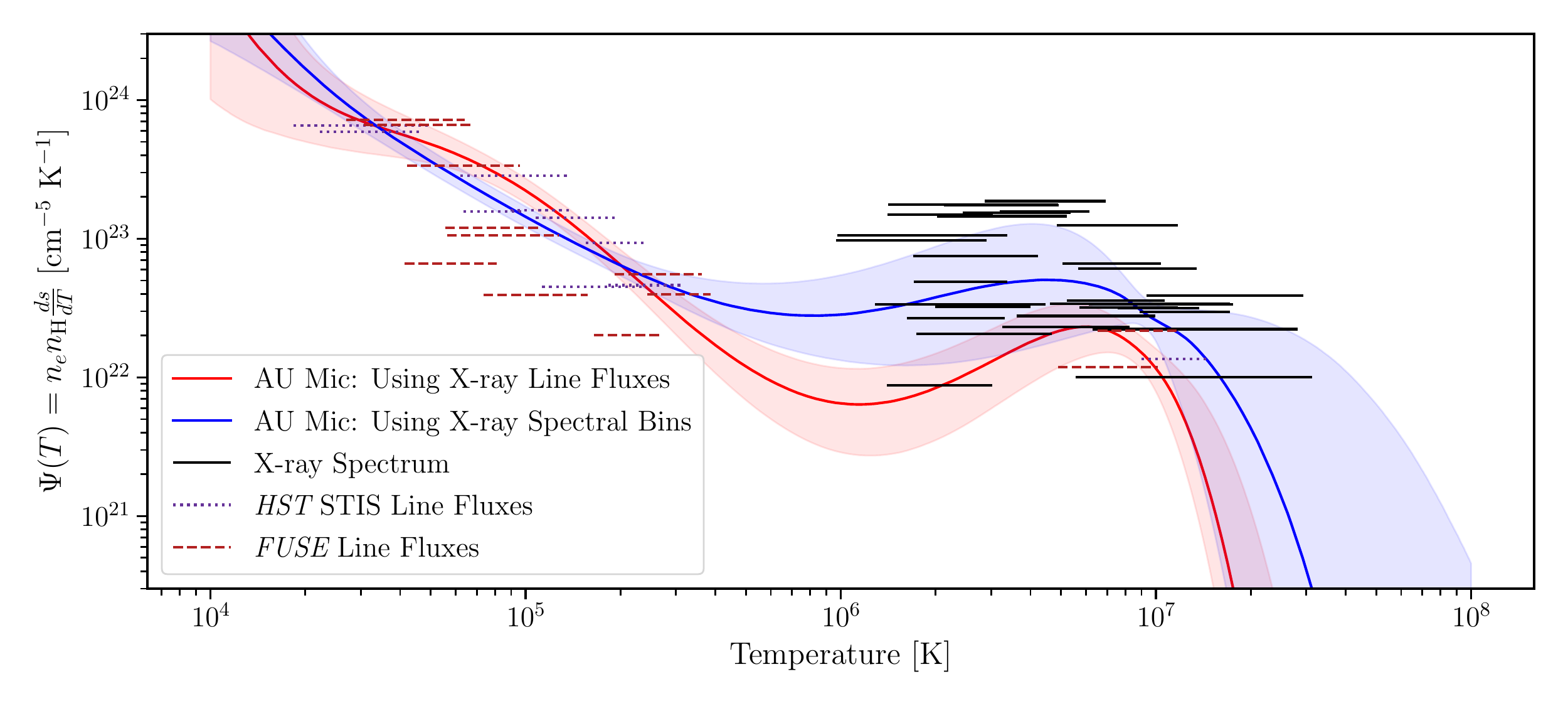}
    \caption{Comparing the DEM of AU Mic fit to a combination of FUV line fluxes from STIS (published in \protect\citealt{Pagano2000}) and \emph{FUSE} (published in \protect\citealt{Redfield2002} and \protect\citealt{Redfield2003}) and either a coarsely sampled X-ray spectrum or a list of X-ray line fluxes reported in \protect\citet{Wood:2018}. The red line and region correspond to the median DEM value and 1$\sigma$ confidence intervals for AU Mic fit using only line flux measurements, while the blue line and region are the model for AU Mic constrained by the X-ray spectrum instead of line fluxes. The horizontal bars represent constraints from the FUV lines listed in Table \ref{table:fuv_lines} and from the \emph{XMM-Newton} X-ray spectrum. The dotted purple bars correspond to the STIS lines published in \protect\cite{Pagano2000} while the dashed red bars correspond to \emph{FUSE} lines published in \protect\citet{Redfield2003} and the black solid bars are from the quiescent \emph{XMM-Newton} spectrum presented in \protect\citet{Kowalski2019AAS} (to be published in Kowalski et al. in prep, and Tristan et al. in prep). There were too many X-ray lines listed in \protect\citet{Wood:2018} to represent in this figure.}
    \label{fig:au_mic_dem}
\end{figure}

In Figure \ref{fig:au_mic_dem} we compare two fits for AU Mic to the constraints imposed by the data, with the red model using only line flux measurements and the blue model using the combination of FUV lines and an X-ray spectrum that will be applicable to fainter M dwarfs. AU Mic has a higher DEM than the Sun across the entire temperature domain, and significantly higher at temperatures greater than $2 \times 10^6$ K, corresponding to the corona. The presence of detectable coronal iron lines, formed at $T > 10^6$ K, places a strong constraint that lifts AU Mic's DEM far higher than the Sun's which had many X-ray spectrum points depressing the DEM in the vicinity of this temperature. The iron emission lines are not solely responsible for the differences, as the constraints from the X-ray spectra set a slope at these high temperatures leading toward the iron lines. The shape of AU Mic's DEM beyond $T = 2 \times 10^5$ K demonstrates the importance of including a corona in calculating the total EUV flux. \citet{Peacock2019b} finds significant differences in the total flux and spectral shape between the \texttt{PHOENIX} models without a corona and the semi-empirical SRPM model from \citet{Fontenla2016}. \citet{Peacock2019b} also simulates the addition of a corona to their models by using the DEM of AU Mic available in \texttt{CHIANTI} from \citet{DelZanna2002}, showing potential opportunities for supplementing stellar atmosphere models with DEMs fit to observations of specific stars.

\begin{figure}
    \centering
    \includegraphics[width=\textwidth]{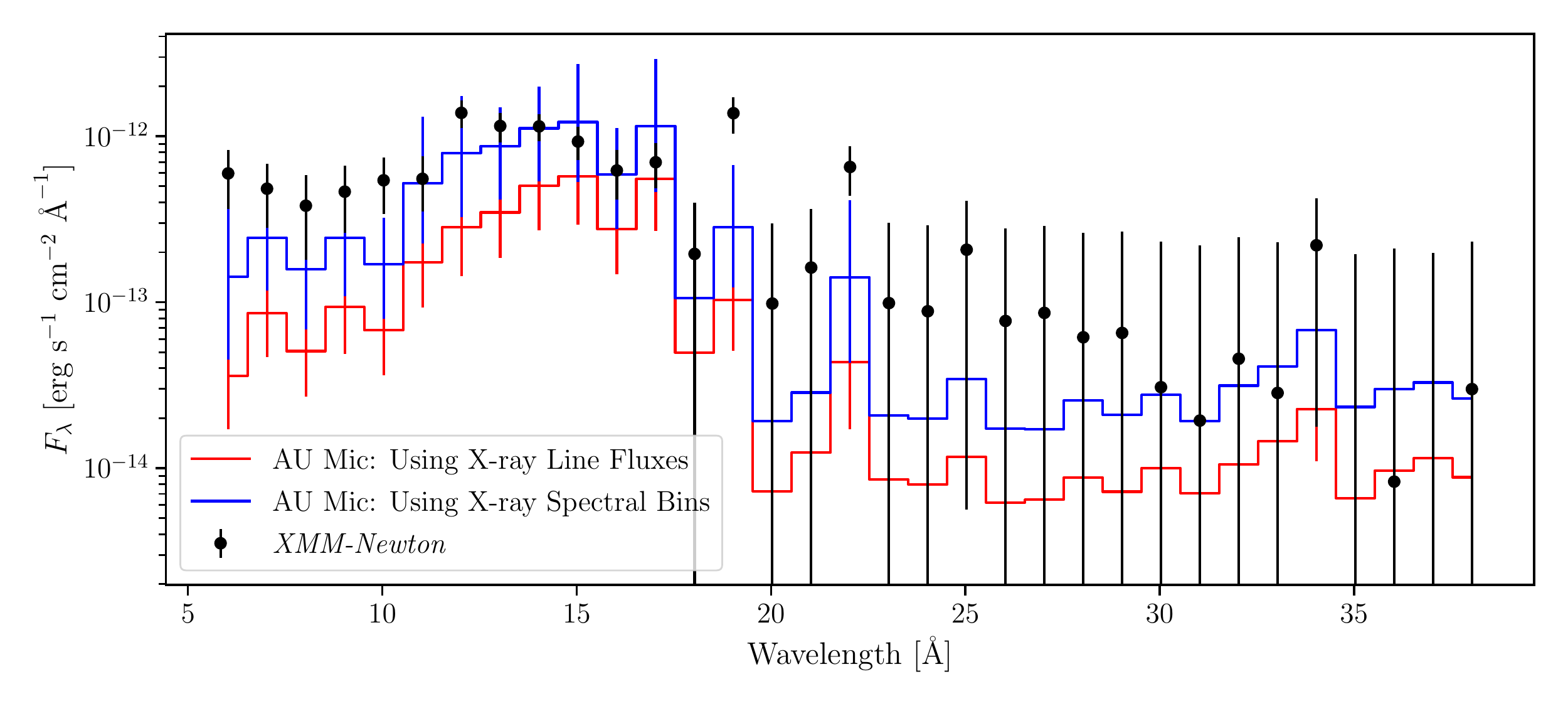}
    \caption{The X-ray data used to fit the DEM compared to the predicted spectra for the DEM models of AU Mic shown in Figure \ref{fig:au_mic_dem}. The black points represent the downsampled RGS X-ray spectrum at a 1 $\textrm{\AA}$ wavelength resolution while the red and blue lines show the DEM prediction for the flux density in the same wavelength bins, incorporating the $s-$factor uncertainty in their error bars, for the models fit with X-ray lines and the X-ray spectrum respectively. They are consistent with each other, although the median prediction of the DEM fit to the spectrum is consistently above the DEM fit to X-ray line fluxes.}
    \label{fig:au_mic_xray}
\end{figure}

\begin{sidewaysfigure}
    \centering
    \includegraphics[width=1.0\textwidth]{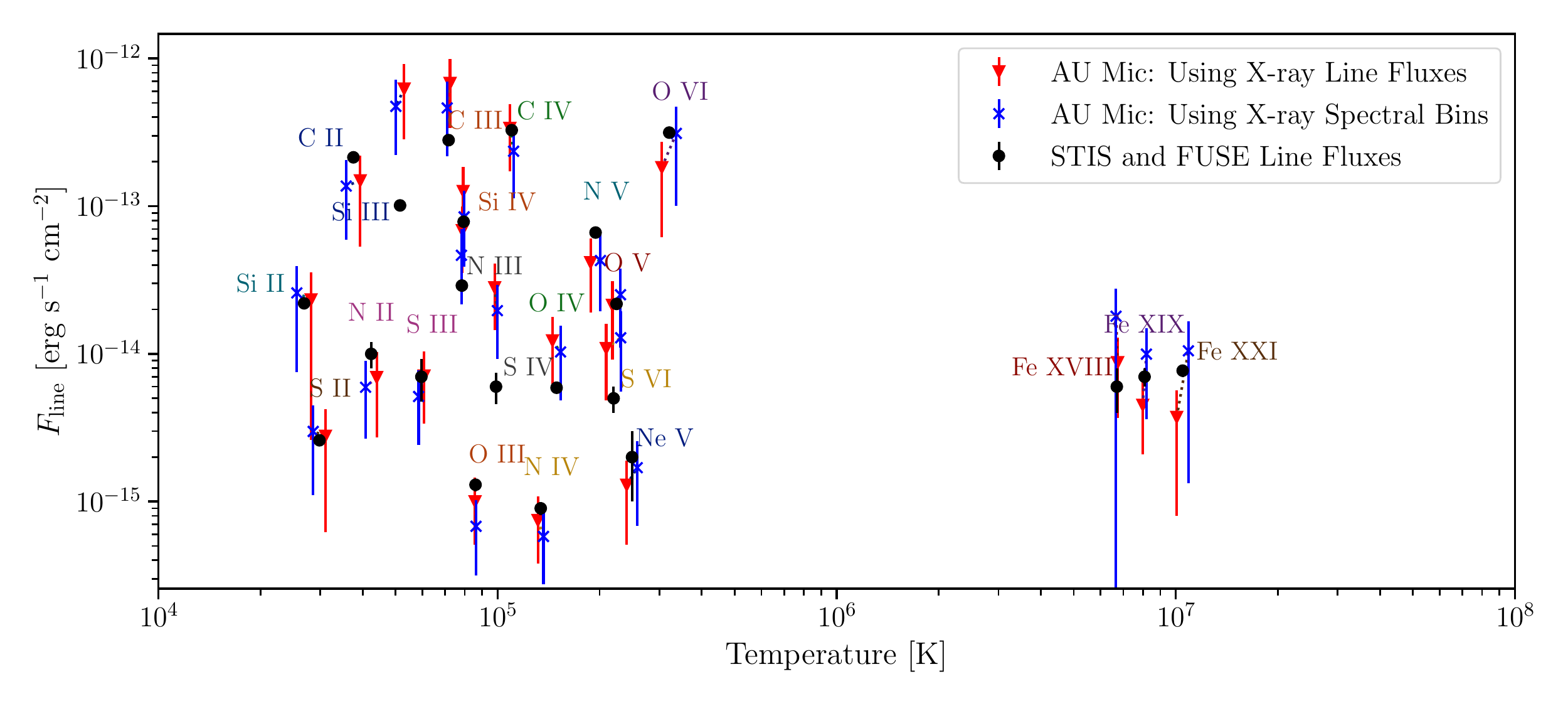}
    \caption{The FUV line fluxes predicted by the AU Mic DEM models shown in Figure \ref{fig:au_mic_dem} compared to the data from \protect\citet{Pagano2000} and \protect\citet{Redfield2003}. Following the color scheme of Figures \ref{fig:au_mic_dem} and \ref{fig:au_mic_xray}, the red triangles represent the predictions from the DEM model fit using only line fluxes while the blue pentagons represent the predictions from the DEM model using the X-ray spectrum. The black points represent the combined fluxes of lines emitted by the ion annotating the point, with errorbars for the measurement uncertainty while the model predictions have errorbars representing the $s-$factor model intrinsic uncertainty.}
    \label{fig:au_mic_lines}
\end{sidewaysfigure}

\begin{figure}
    \centering
    \includegraphics[width=\textwidth]{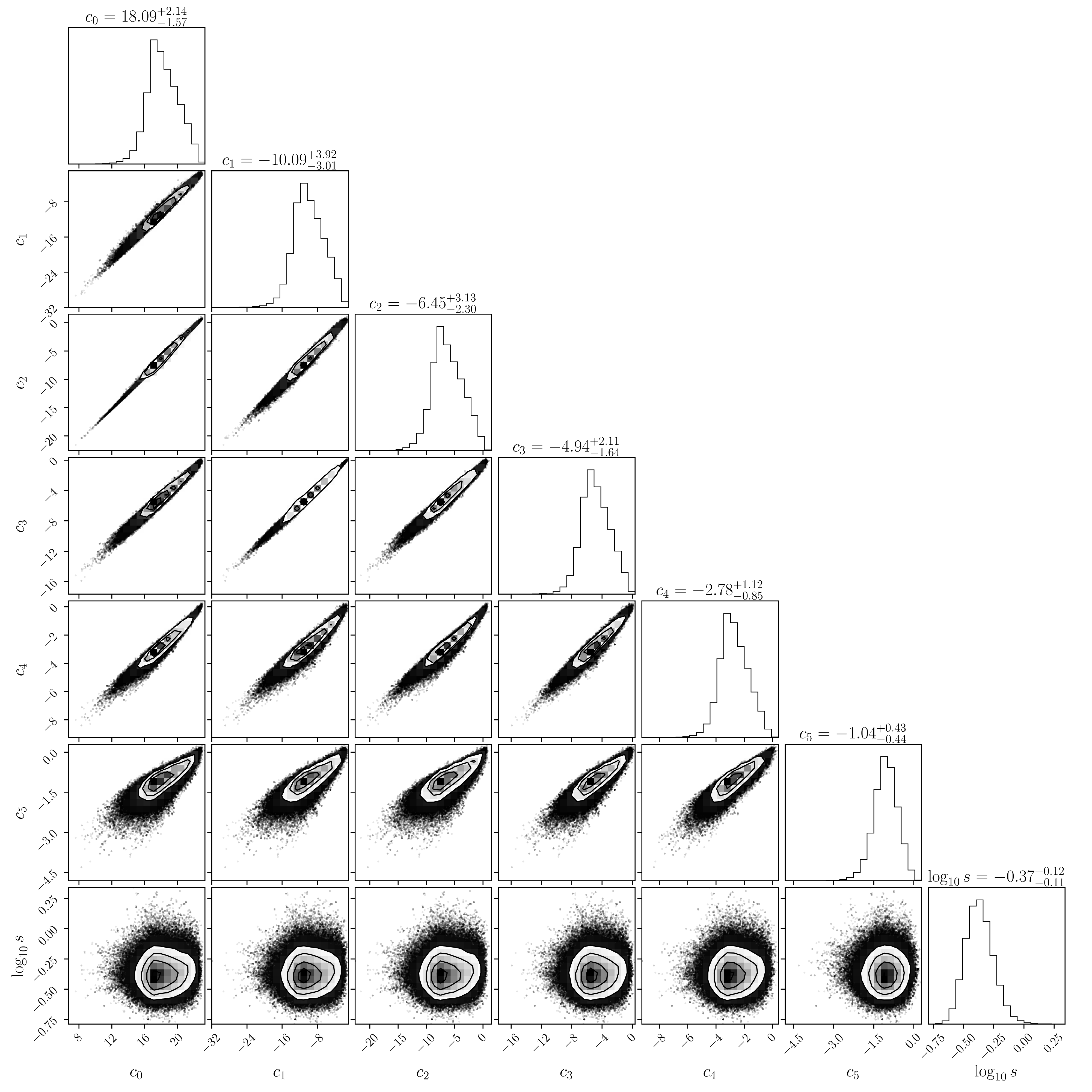}
    \caption{The corner plot showing the parameter distributions when fitting AU Mic with a 5$^{\textrm{th}}$ order polynomial to FUV lines and the X-ray spectrum. The model uncertainty for AU Mic is slightly better than that of the Sun, with $s = 0.4$ instead of 0.6.}
    \label{fig:au_mic_line_fit_corner}
\end{figure}

Figure \ref{fig:au_mic_xray} compares the predicted X-ray spectra to the observed spectrum and Figure \ref{fig:au_mic_lines} compares the predicted line fluxes to the FUV line profile measurements listed in Table \ref{table:fuv_lines}. Both figures incorporate the $s-$factor uncertainties and DEM shape variation (Section \S\ref{sec:euv_fitting}) and show that the model predictions are typically consistent with the data to 1$\sigma$. Figure \ref{fig:au_mic_line_fit_corner} shows the parameter distributions for the DEM fit to line fluxes, and we find that the $s-$factor for AU Mic is 0.4, comparable to that of the Sun. At lower wavelengths the red model, which was not fit to the spectrum itself, significantly underestimates the flux. This may be a consequence of the higher energy emission including flux from free-free or free-bound continuum sources, creating the discrepancies between both these DEMs. To reproduce the flux in these bins without accounting for this extra emissivity, the DEM fit to the spectrum must enhance the amount of material at these high temperatures. Including free-free and free-bound continuum emissivities in the $G_{\lambda} (T)$ matrices should mitigate or eliminate these discrepancies.

\begin{deluxetable}{cccccc}

\tablecaption{Ion fluxes of AU Mic compared to the predictions from the red DEM model shown in Figure \ref{fig:au_mic_dem}.\label{table:fuv_lines}}
\tablehead{
\colhead{Ion} & \colhead{Wavelengths} & \colhead{$\log_{10} T_f$} & \colhead{Observed Flux}  & \colhead{DEM Prediction}\\
\colhead{}
& \colhead{[$\textrm{\AA}$]}
& \colhead{$\log_{10}(\textrm{[K]})$} 
&\colhead{[$10^{-15}$ erg\,s$^{-1}$\,cm$^{-2}$]}
& \colhead{[$10^{-15}$ erg\,s$^{-1}$\,cm$^{-2}$]}}
\startdata
\ion{C}{2}   & 1324, 1336\tablenotemark{a} & 4.6   & $214 \pm 16.0$   & $149^{+71.7}_{-96.2}$\\
\ion{C}{3}   & 977, 1176\tablenotemark{a} & 4.9   & $280 \pm 16.0$   & $681^{+309}_{344}$\\
\ion{C}{4}   & 1548, 1551 & 5.0   & $327 \pm 24.2$   & $208^{+164}_{-174}$\\
\ion{N}{2}   & 1086 & 4.6   & $10.0 \pm 2.00$  & $6.94^{+3.33}_{-4.21}$\\
\ion{N}{3}   & 990, 992 & 4.9   & $29.0 \pm 2.24$  & $68.6^{+30.9}_{-33.3}$\\
\ion{N}{4}   & 1487 & 5.1   & $0.90 \pm 0.09$  & $0.747^{+0.337}_{-0.366}$\\
\ion{N}{5}   & 1238.8, 1242.8 & 5.3   & $66.3 \pm 4.99$  & $41.6^{+19.1}_{-22.5}$\\
\ion{O}{3}   & 1666 & 4.9   & $1.30 \pm 0.13$  & $1.00^{+0.448}_{-0.489}$\\
\ion{O}{4}   & 1400, 1401, 1407 & 5.1   & $5.90 \pm 0.423$  & $12.3^{+5.55}_{-6.17}$\\
\ion{O}{5}   & 1218, 1371 & 5.3   & $21.8 \pm 1.86$  & $21.4^{+9.84}_{-12.2}$ \\
\ion{O}{6}   & 1032, 1038 & 5.5   & $315 \pm 26.4$  & $183^{+89.7}_{-121}$ \\
\ion{Ne}{5}   & 1146 & 5.4   & $2.00 \pm 1.00$  & $1.30^{+0.599}_{-0.784}$\\
\ion{Si}{2}  & 1260, 1265, 1304, 1527, 1533 & 4.3   & $22.0 \pm 1.06$  & $23.3^{+12.3}_{-20.7}$\\
\ion{Si}{3}  & 1108, 1110, 1113, 1206, 1295, 1297, 1299, 1301, 1303 & 4.8   & $101 \pm 8.11$  & $624^{+289}_{-340}$\\
\ion{Si}{4}  & 1394, 1403 & 4.9   & $78.5 \pm 5.73$  & $126^{+57.3}_{-62.8}$\\
\ion{S}{2}   & 1253.8, 1259.5 & 4.4   & $2.60  \pm 0.18$ & $2.78^{+1.45}_{-2.16}$\\
\ion{S}{3}   & 1016, 1021,  & 4.8   & $7.00  \pm 2.24$ & $7.11^{+3.29}_{-3.75}$\\
\ion{S}{4}   & 1063, 1073 & 5.0   & $6.00  \pm 1.41$ & $28.2^{+12.6}_{-13.7}$\\
\ion{Fe}{18} & 975 & 6.9   & $6.00 \pm 2.00$  & $8.75^{+4.10}_{-5.08}$\\
\ion{Fe}{19} & 1118 & 7.0   & $7.00 \pm 1.00$  & $4.51^{+2.05}_{-2.42}$\\
\ion{Fe}{21}& 1354 & 7.0   & $7.70 \pm 0.77$   & $3.73^{+1.91}_{-2.93}$\\
\enddata
\tablecomments{All entries are from STIS line fluxes reported in \citet{Pagano2000} and/or \emph{FUSE} measurements reported in \citet{Redfield2002} and \citet{Redfield2003}.}
\tablenotetext{a}{multiplet}
\end{deluxetable}

\section{Our DEMs Compared to Published Literature}\label{sec:literature_comparison}
\begin{figure}
    \centering
    \begin{minipage}{.48\textwidth}
        \includegraphics[width=\textwidth]{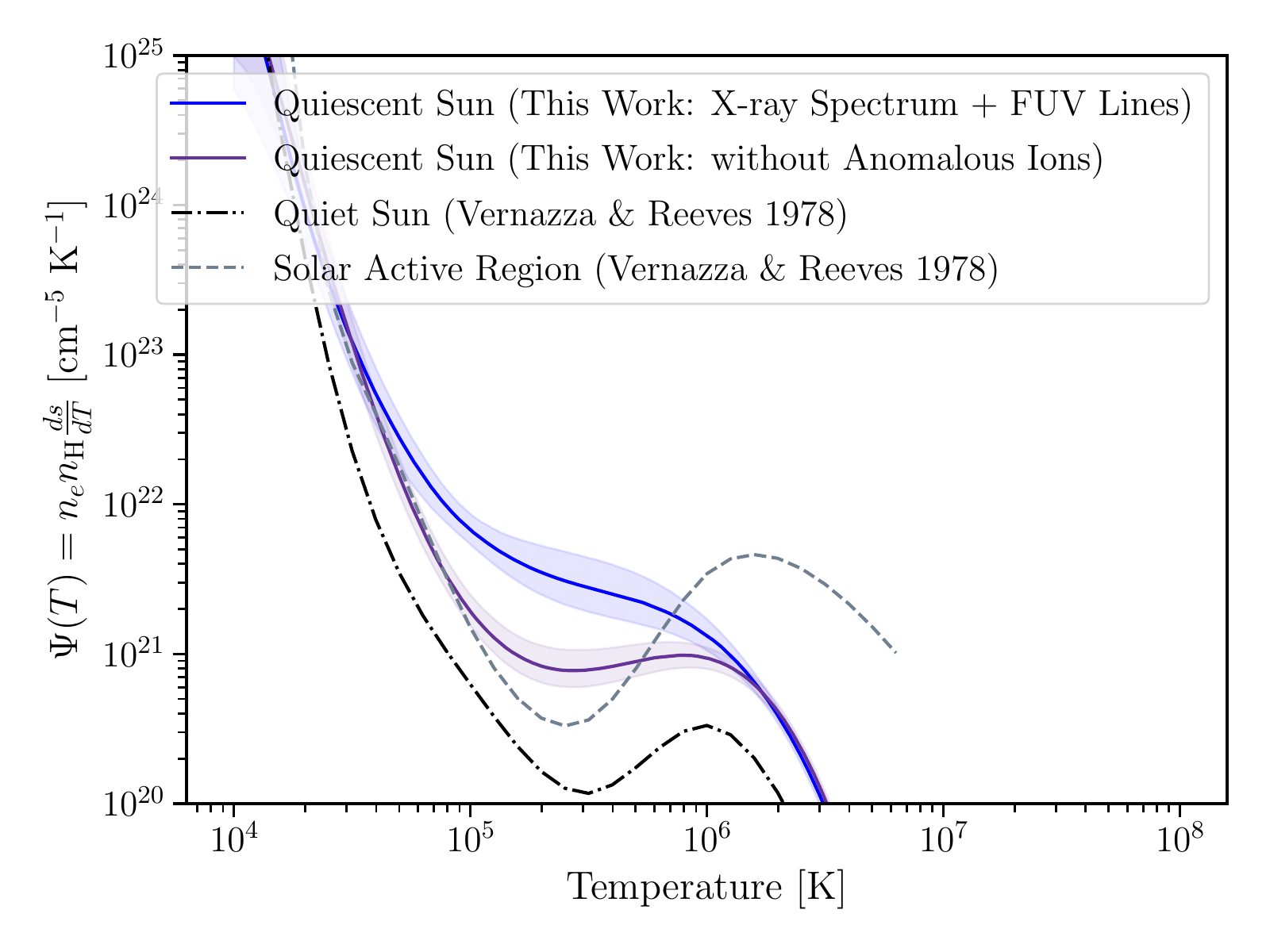}
    \end{minipage}
    \begin{minipage}{.48\textwidth}
        \includegraphics[width=\textwidth]{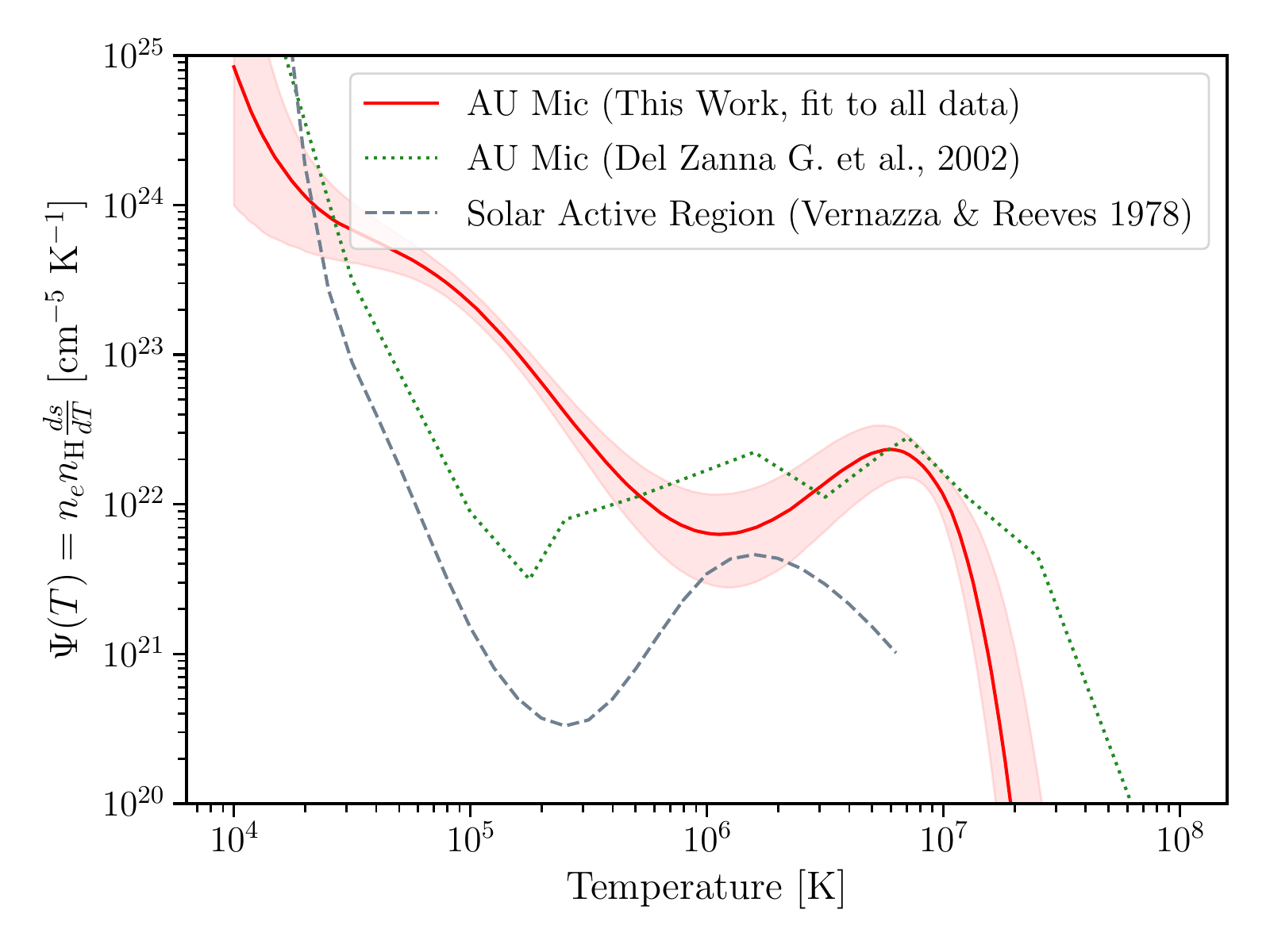}
    \end{minipage}
    \caption{The left panel compares both of our DEMs for the quiescent disk-integrated Sun described in Section \S\ref{sec:euv_fitting} to the ``quiet Sun" DEM derived by \protect\citet{Vernazza1978} from observations of quiet regions of the Sun, while the right panel compares our DEM for AU Mic to the AU Mic DEM published by \protect\citet{DelZanna2002} and the DEM of an active region published by \protect\citet{Vernazza1978}. These DEMs used for comparison were available in the CHIANTI database \protect\citep{Dere1997, DelZanna2015}.}
    \label{fig:lit_dem}
\end{figure}

Differential emission measure techniques are primarily applied in the solar context to resolved regions, highlighting individual structures like coronal holes or flare loops. \citet{Vernazza1978} fit DEMs to many such structures, including ``quiet regions" with minimal observed activity during the 9-month interval from 1973 May to 1974 August, shortly before the solar minimum of Cycle 21. This is not perfectly analogous to our quiescent Sun DEM fit to data integrated over the entire solar disk, but it is the best comparison for which we had access to a published DEM via \texttt{CHIANTI} \citep{Dere1997, DelZanna2015}. The left panel of Figure \ref{fig:lit_dem} compares the \citet{Vernazza1978} quiet region DEM to our disk-integrated quiescent Sun DEM, fit under the assumptions applicable to observing the Sun as a star (albeit with much higher signal-to-noise): the few FUV lines listed in Table \ref{table:sun_lines} and the X-ray spectrum at a low resolution of $R \leq 50$. We also include the solar DEM fit excluding anomalous ions and including EUV lines.

Similarly, the right panel of Figure \ref{fig:lit_dem} compares our DEM for AU Mic to the DEM published by \citet{DelZanna2002}, which combined \emph{FUSE}, STIS, and \emph{EUVE} observations. \texttt{CHIANTI} has a volume differential emission measure version of the \citet{DelZanna2002} DEM which needed to be divided by $\frac{4\pi R_{\star}^2}{d^2}$ to match our formulation of a column differential emission measure, and we use the stellar radius and distance assumed by \citet{DelZanna2002}, $R_{\star} = 0.68 R_{\odot}$ and $d = 9.94$ pc, for consistency in the scaling factor. Like our dataset, they did not have access to simultaneous observations from these different instruments. Unlike our dataset, they included the integrated fluxes of lines observed with \emph{EUVE}, but we believe the \emph{EUVE} observations they used were flare contaminated for reasons discussed later in Section \S\ref{sec:final_model_spectra}. We also compare our AU Mic DEM to the active solar region DEM of \citet{Vernazza1978}, demonstrating a small resemblance between the active region and the active star AU Mic. The active region DEM seems to shift the shape of the quiet sun DEM to a higher temperature and dramatically enhance the DEM near $10^6$ K. AU Mic, a star more active than the Sun, has more material at nearly all temperatures.

Both panels show that our DEMs are significantly higher than their literature counterparts at temperatures near $10^5$ K, roughly corresponding to the transition region. For the AU Mic panel on the right, the \citet{DelZanna2002} DEM is significantly lower than ours between $3 \times 10^4$ and $3 \times 10^5$ K. \citet{Monsignori1996} published the time-evolution of the AU Mic DEM during a flare observed by \emph{EUVE} in July 1992, and while they use a slightly different formulation of the DEM $= n_{e}^2 \frac{dV}{dT}$, the shape of our DEM strongly resembles theirs published in panels a and g of their Figure 9, which correspond to quiescent phases, while multiplying our DEM by $4\pi R_{\star}^2 \approx 3\times 10^{22}$ cm$^2$ scales to the approximately the same order of magnitude as their DEM. Unfortunately, their DEM was not available in \texttt{CHIANTI} for direct comparison.

The discrepancy at temperatures below $10^6$ K is largely driven by the line list we are using to fit the DEM, where our strongest lines are from Na-like and Li-like species \ion{N}{5}, \ion{Si}{4}, and \ion{C}{4}, and the solar DEM excluding these ions is much closer to the \citet{Vernazza1978} quiet region DEM. \citet{DelZanna2002} fits the DEM of AU Mic without these lines and has enough individual lines to constrain the DEM in this temperature regime without them. We will not be able to afford this luxury for nearly every other M dwarf unless we are observing a strong flare. Instead we fit our DEM including these lines and use the $s-$parameter boost to our variance to account for the systematic uncertainties involved. As long as some other ions formed in the same region are included in the line list, even with just upper limits on their fluxes, the DEM will shift down to accommodate these lines. In the future, we plan to test dividing the observed fluxes of these lines by a corrective factor $\sim 5$ before fitting the DEM to see how this improves the fit and affects the predicted EUV spectrum.

Working in a data-limited regime is also why we fit for the DEM using an assumed functional form instead of interpolating between the emission measure loci estimated from individual lines \citep{Pagano2000}, or fitting for the value of the DEM in discrete temperature bins \citep{DelZanna2002, DelZanna2015}. Without measured lines in the temperature regime corresponding to most EUV lines, we must use our assumption of a continuous function anchored on both ends of the inaccessible temperature/wavelength regime. The DEM for AU Mic derived by \citet{DelZanna2002} is poorly constrained between $10^{6} < T < 10^{6.7} K$ because they do not use X-ray line fluxes or a coarse spectrum, but our DEMs agree on the position and magnitude of the coronal peak DEM ($T=10^7$ K), if not the shape of the decline. Other contributions to the discrepancies between DEMs are differences in the atomic data and calculations for abundances and the ionization equilibrium and level populations. While the DEMs created by \citet{SanzForcada2011} were not available in \texttt{CHIANTI}, we expect significant discrepancies at the lower temperature end of the DEM because the majority of their cool dwarfs did not have UV data available.

\section{Comparing Model EUV Spectra to Data}\label{sec:final_model_spectra}
\begin{figure}
    \centering
    \includegraphics[width=\textwidth]{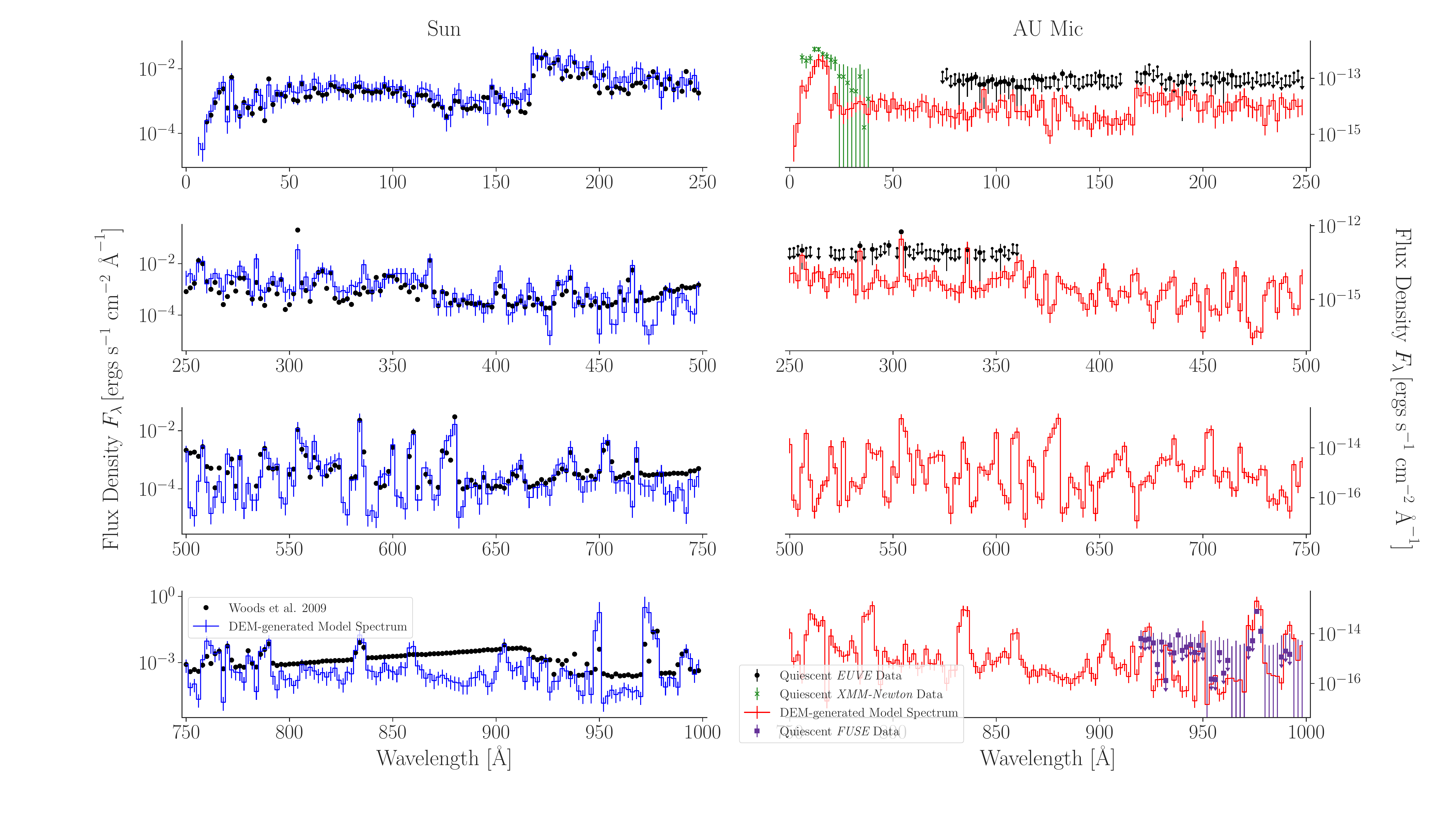}
    \caption{The individual rows of this figure show 250 $\textrm{\AA}$ chunks of the X-ray, extreme ultraviolet, and a portion of the far ultraviolet regimes. The left side compares to the \protect\citet{Woods2009} spectrum shown as black points to the DEM-generated model spectrum for the Sun in blue, where the DEM was fit to EUV lines, FUV lines excluding anomalous ions, and X-ray data. The right side shows the DEM-generated model spectrum for AU Mic in red, compared to both \emph{EUVE} data shown as black points and \emph{FUSE} data shown as purple squares. The model spectrum for the Sun is largely consistent with the data, except for a recombination continuum associated with the \ion{H}{1} 912 $\textrm{\AA}$ line spanning $\sim 800$ to $912$ $\textrm{\AA}$. The AU Mic \emph{EUVE} data is almost entirely consistent with zero, but our model does not predict flux significantly higher than these upper limits, while roughly matching some of the clearly detected emission lines.}
    \label{fig:euv_panel}
\end{figure}

With our implementation of the DEM well-characterized, we can move on to the main objective of this project: generating EUV spectra in a format useful to the astronomical community with errorbars that self-consistently account for both statistical and systematic uncertainties. Our model spectra range from 1 to 2000 $\textrm{\AA}$ at a constant $R= \frac{\Delta \lambda}{\lambda} = 500$, but we advise using data instead of our model in the regimes where that is possible. The model spectrum files include uncertainties derived according to the method described in Section \S\ref{sec:euv_fitting}. Figure \ref{fig:euv_panel} compares our model spectrum of the Sun to the \citet{Woods2009} data and our model spectrum of AU Mic to an \emph{EUVE} observation from July 1992 and quiescent \emph{FUSE} data from a different time described in \citet{Redfield2002, Redfield2003}. This \emph{EUVE} observation was during the calibration phase of the mission and happened to catch a flare, first reported by \citet{Cully1993} and later studied in more detail in \citet{Monsignori1996}. We compare our DEM-generated spectrum to data from the quiescent time segment before the flare, extracted using standard \emph{EUVE} Guest Observer Center IRAF procedures. Useful data were obtained with the short wavelength (70 to 190 $\textrm{\AA}$) and medium wavelength (140 to 380 $\textrm{\AA}$) detectors, but no useful signal was present in the long wavelength (280 to 760 $\textrm{\AA}$) region. The spectral resolution of the 3 spectrometers is $\sim$ 0.5, 1.0, and 2 $\textrm{\AA}$ , which corresponds to 7 pixels per resolution element on the detectors. The photon event data were screened to eliminate high background times and times when the detectors were switched off. The IRAF routine ``\texttt{apall}" was used for the spectral extraction with a 14 pixel wide spectral region and two 85 pixel wide background regions measured above and below the stellar spectrum. The count rate spectra were converted to flux densities using the effective areas established by the \emph{EUVE} mission and the oversampled spectra were smoothed to the intrinsic spectrometer resolution. The large background area sampled allowed precise monitoring of the time dependent background. \citet{DelZanna2002} used a 1993 \emph{EUVE} observation which has no published lightcurve to verify the absence of flares and the time-averaged 1993 spectrum shows more flux than the quiescent 1992 data at all wavelengths indicating possible flare contamination.

Our DEM prediction for the EUV luminosity of AU Mic is $L_{\mathrm{EUV}} = 9.0^{+6.1}_{-4.6} \times 10^{28}$ erg s$^{-1}$, while the \citet{Linsky2014} relations give $1.2 \times 10^{29}$ erg s$^{-1}$,  \citet{SanzForcada2011} relation gives $1.4 \times 10^{30}$ erg s$^{-1}$, and \citet{Chadney2015} finds $L_{\mathrm{EUV}} = 8.4 \times 10^{28}$ erg s$^{-1}$ using a coronal emission measure distribution model. We use the Lyman-$\alpha$ flux reported by \citet{Wood2005} for the \citet{Linsky2014} relations, and the X-ray luminosity obtained by multiplying the integrated flux of our EPIC-MOS spectrum with $4 \pi d^2$, $L_{\textrm{X-ray}} = 3.0 \times 10^{29}$ erg s$^{-1}$, for the \citet{SanzForcada2011} relation. To convert the surface EUV flux reported by \citet{Chadney2015} to luminosity we use their assumed radius for AU Mic $= 0.68 R_{\odot}$. The major advantages of our DEM approach are a well-characterized uncertainty and a balance between ease of implementation and specificity to each star. Using Equation 3 of \citet{SanzForcada2011} has a minimum uncertainty of 1.99 dex in predicting the EUV flux, but our method can do significantly better than this empirical relation even for very faint targets like TRAPPIST-1, with only a few FUV line measurements and an integrated X-ray flux or coarse spectrum. The DEM method also provides a low-resolution spectral shape in addition to a total flux, which may be useful for those who wish to model more detailed effects of high energy stellar radiation on a planet, disk, or the local interstellar medium.

\section{Case Studies}\label{sec:case_studies}
Thus far, we have demonstrated our method on targets with extremely good data, which are not representative of the majority of stars for which the astronomical community needs reconstructed EUV spectra to enable other science. In this section we apply our method to three M dwarfs of interest to the exoplanet community, all fainter and less active than AU Mic, and show that their coronae are hotter than the Sun and will seriously affect the atmospheric evolution of planets orbiting in their respective habitable zones.

\subsection{GJ 832}\label{sec:gj832}
\begin{figure}
    \centering
    \includegraphics[width=0.5\textwidth]{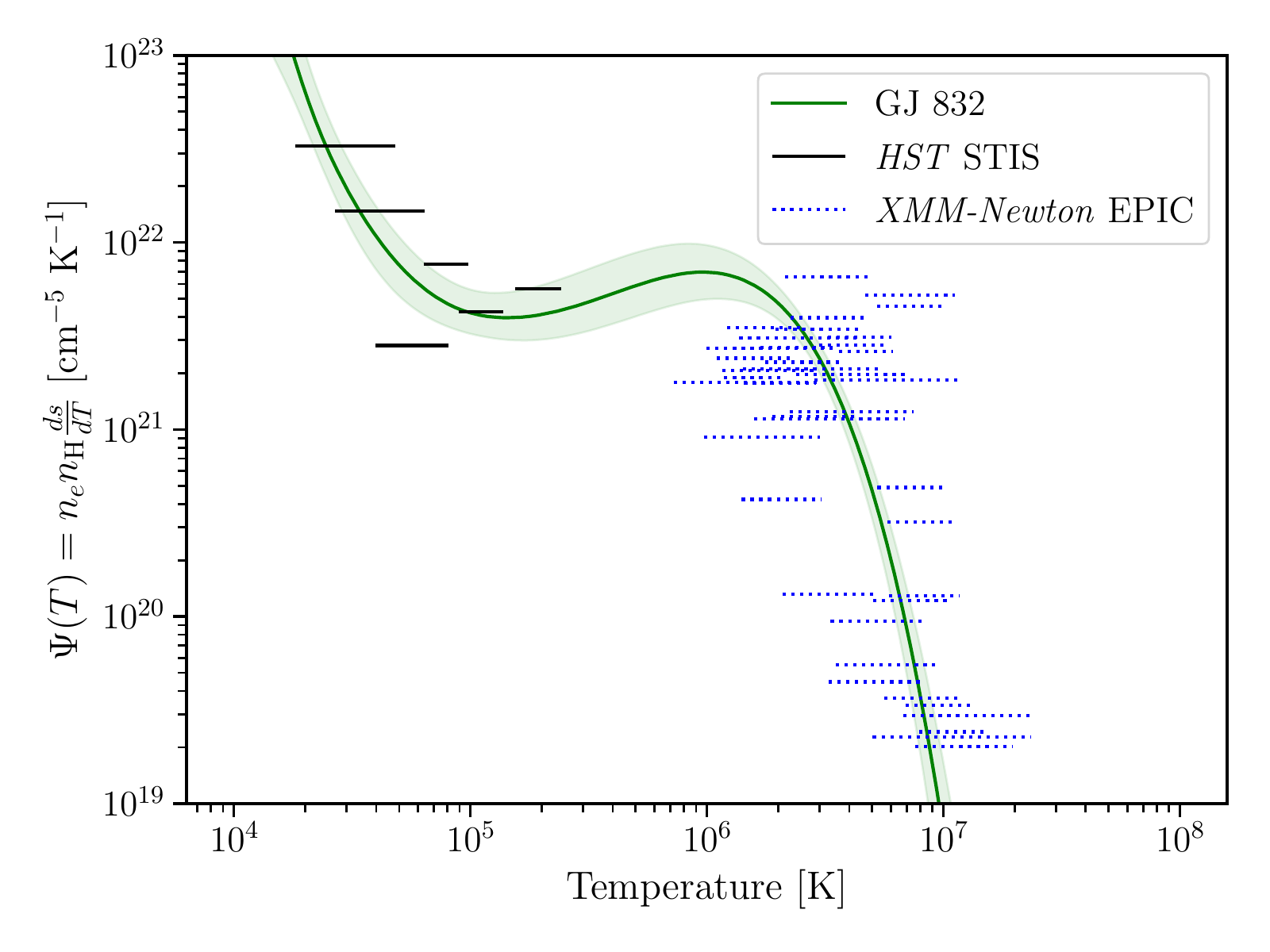}
    \caption{Comparing the DEM of GJ 832 fit to a quiescent X-ray spectrum from \emph{XMM-Newton} EPIC and line fluxes from STIS, with $\overline{\Psi}(T)$ constraints shown as horizontal dotted blue and solid black bars respectively.}
    \label{fig:gj832_dem}
\end{figure}

\begin{figure}
    \centering
    \includegraphics[width=\textwidth]{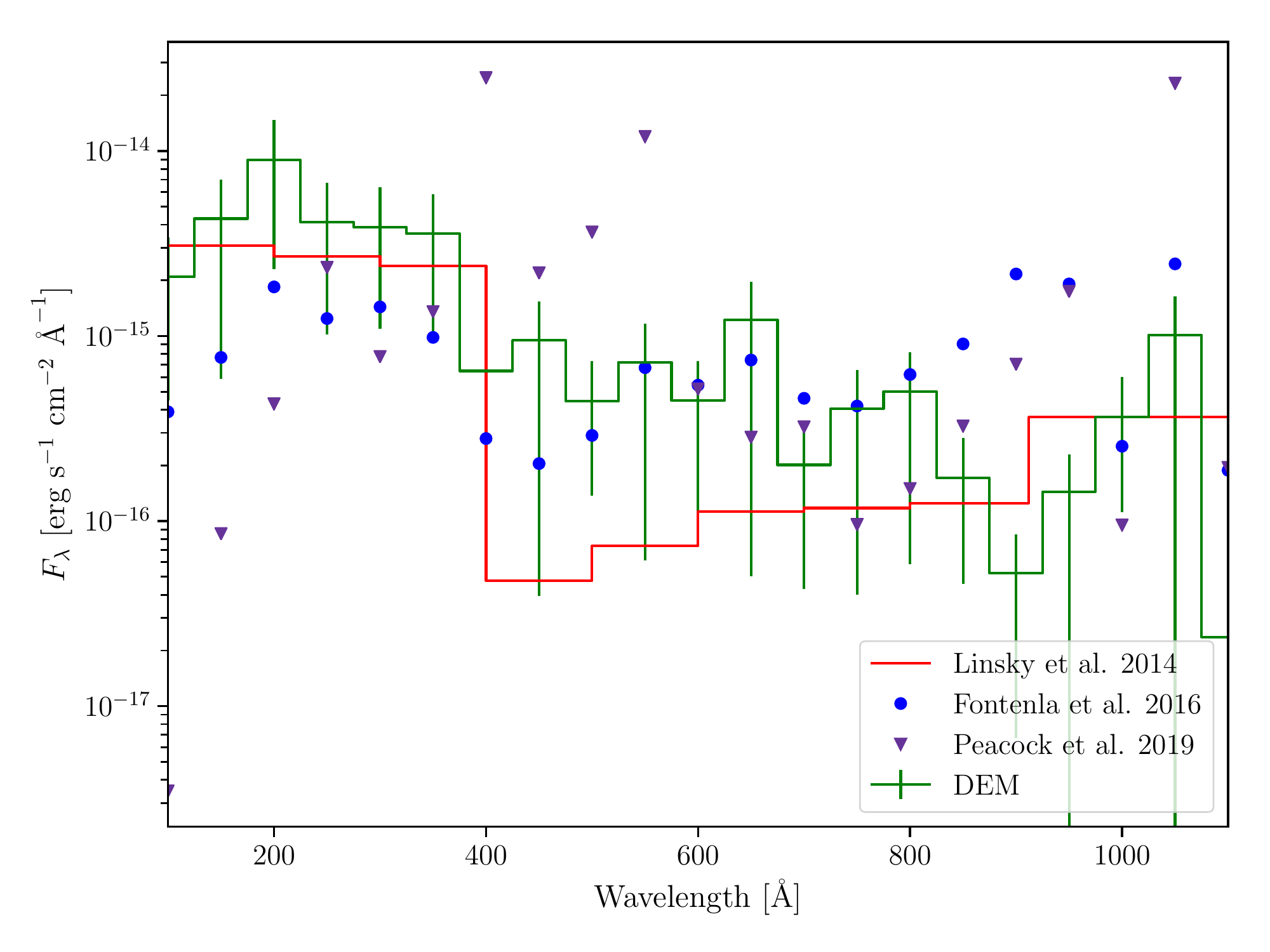}
    \caption{Comparing four methods of reconstructing the EUV spectrum of GJ 832. The green line and errorbars represent the spectrum predicted by our DEM model with the errors enclosing the $16^{\textrm{th}}$ to $84^{\textrm{th}}$ percentile intervals for the predicted flux according to the method described in Section \S\ref{sec:euv_fitting}. The red lines show the predicted EUV flux in 100 $\textrm{\AA}$ bandpasses according to the Lyman-$\alpha$ correlations published in \protect\citet{Linsky2014}, available in the MUSCLES dataset. The blue points represent the semi-empirical model published by \protect\citet{Fontenla2016}, also available in the MUSCLES dataset. The purple triangles show the \texttt{PHOENIX} model published in \protect\citet{Peacock2019b} and hosted on \emph{MAST}. These four methods of predicting the EUV spectrum appear to have different regions of mutual agreement, discussed in Section \ref{sec:gj832}.}
    \label{fig:gj832_euv}
\end{figure}

GJ 832 was included in the original MUSCLES survey  \citep{France2016} with its EUV flux estimated by the \citet{Linsky2014} correlations and semi-empiricaly modeled by both \citet{Fontenla2016} and \citet{Peacock2019b}. We fit our DEM model to the line fluxes published in \citet{Youngblood2016} and X-ray data hosted on \emph{MAST} as part of the MUSCLES data products. The line fluxes were measured using STIS data while the X-ray spectrum is from \emph{XMM-Newton} EPIC \citep{France2016}. Comparing our model spectra to the semi-empirical models and the EUV fluxes predicted by the Linsky relations in Figure \ref{fig:gj832_euv} shows that the different models agree with each other in different wavelength regimes. The data we used to fit GJ 832 was not perfectly quiescent, so some of the model discrepancy may be due to flare contamination \citep{Youngblood2016}. The DEM inferred from fitting the data is shown in Figure \ref{fig:gj832_dem}.

The DEM model tends to predict higher fluxes than the other three models at wavelengths shorter than 600 $\textrm{\AA}$. Below 400 $\textrm{\AA}$ all models except for the \citet{Peacock2019b} \texttt{PHOENIX} model coincide quite closely, with the outlier lacking a coronal contribution. Between 400 to 600 $\textrm{\AA}$ both the DEM and semi-empirical models predict higher fluxes than the Lyman-$\alpha$ correlations. The biggest discrepancies between models are between 800 to 1100 $\textrm{\AA}$ where the semi-empirical models include the recombination continuum at the \ion{H}{1} 912 $\textrm{\AA}$ line and the blue wing of the Lyman-$\alpha$ 1216 $\textrm{\AA}$ line, both of which are unaccounted for by the DEM model and which the Lyman-$\alpha$ correlation fluxes seem to underestimate. \citet{Tilipman2020} updates the \citet{Fontenla2016} SSRPM model of GJ 832, and when all these EUV reconstruction methods have been applied to a larger sample of stars we may have more insight into the conditions under which each is more likely to be accurate.

\subsection{Barnard's Star}\label{sec:gj699}
\begin{figure}
    \centering
    \begin{minipage}{0.45\textwidth}
        \includegraphics[width=1.0\textwidth]{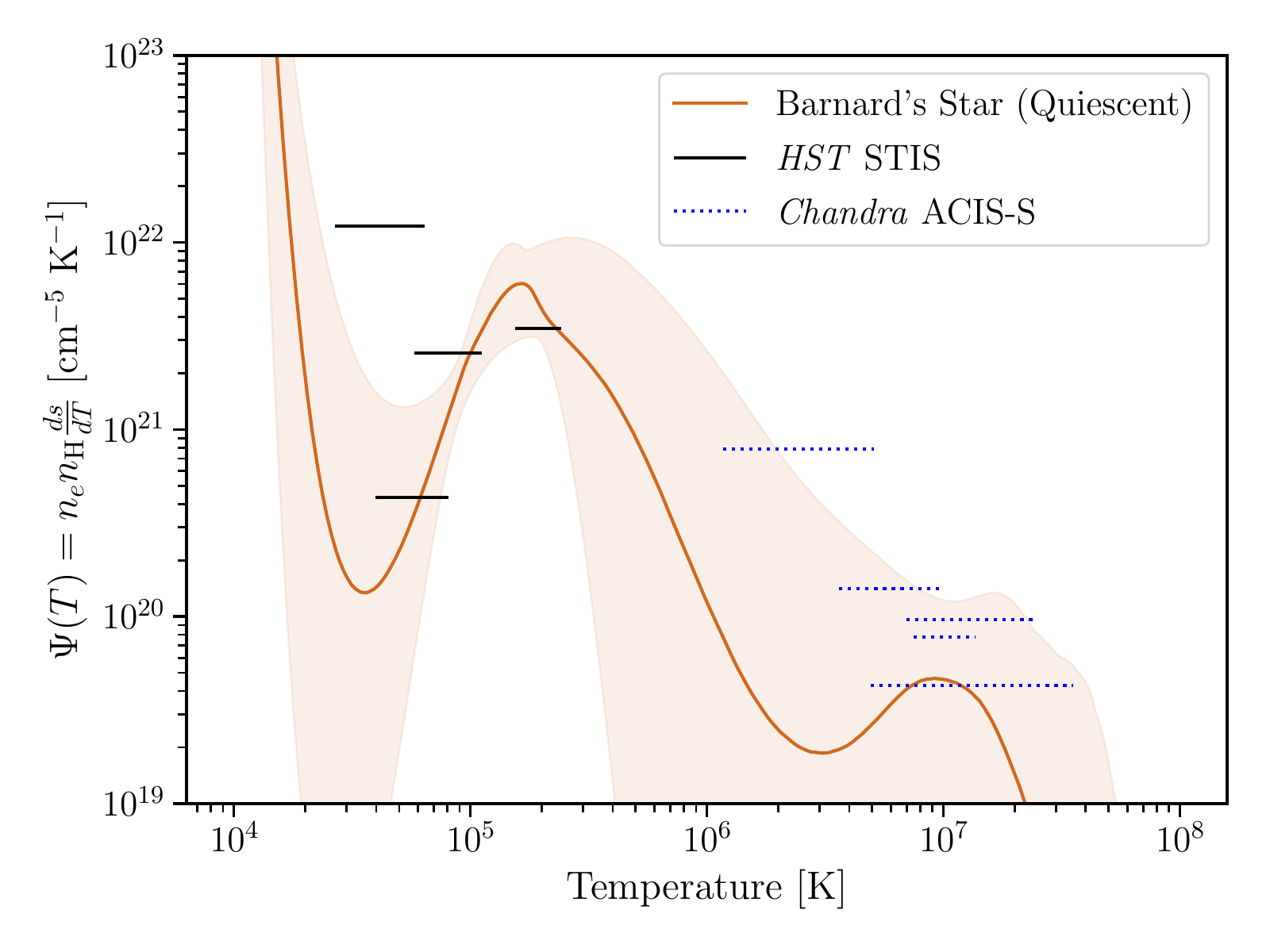}
    \end{minipage}
    \begin{minipage}{0.45\textwidth}
        \includegraphics[width=1.0\textwidth]{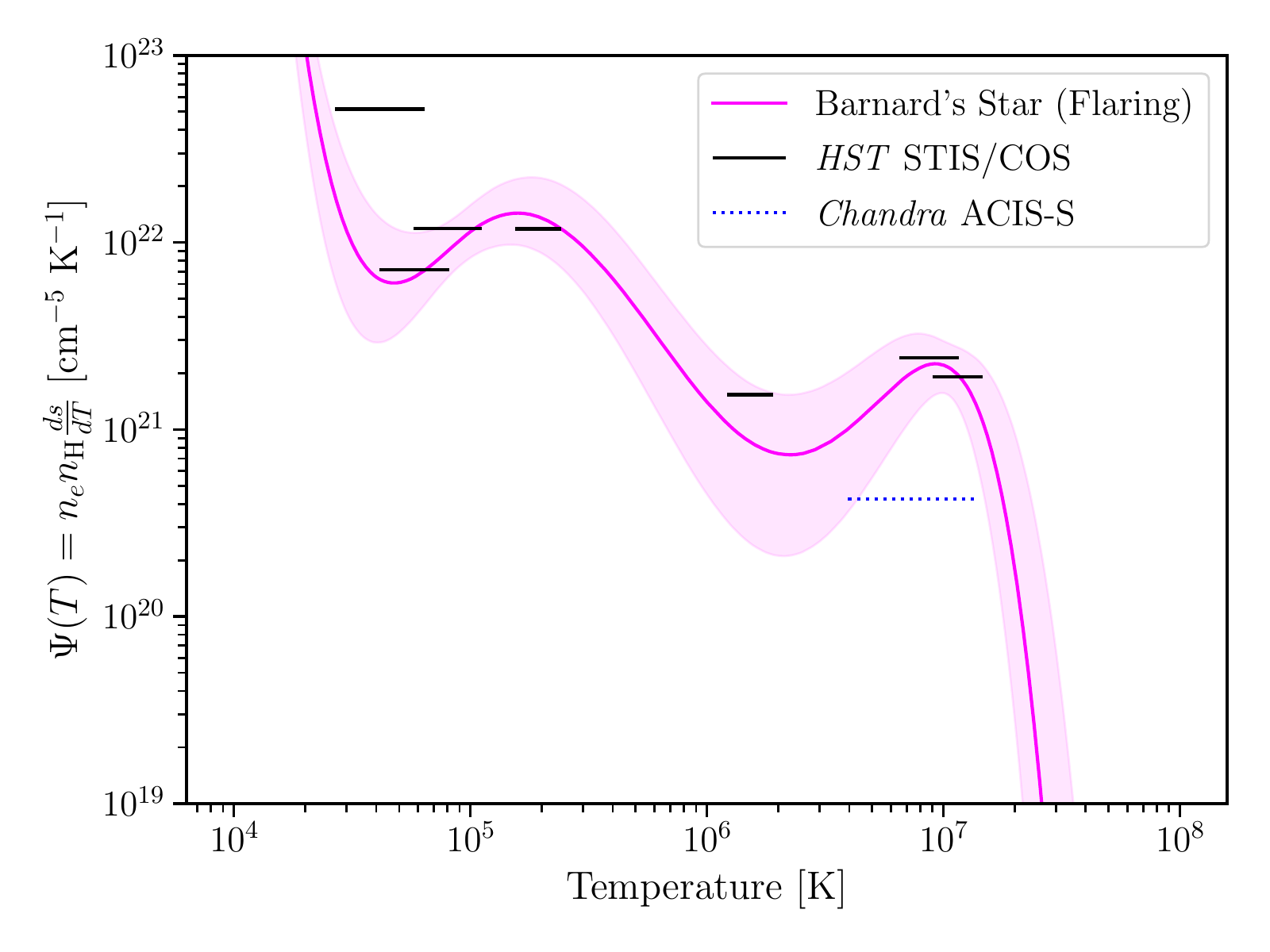}
    \end{minipage}
        \caption{The left panel compares the DEM of Barnard's Star fit to a coarse X-ray spectrum from \emph{Chandra} ACIS-S and line fluxes measured from \emph{HST} STIS. The right panel compares the DEM of Barnard's Star fit to an integrated X-ray flux from \emph{Chandra} ACIS-S and line fluxes from STIS/COS during a flare. For both panels, the dotted blue and solid black bars represent the $\overline{\Psi}(T)$ constraints for the X-ray and FUV data respectively.}
    \label{fig:gj699_dem}
\end{figure}

Barnard's Star is old and inactive compared to most M dwarfs \citep{Ribas2018}, but it still flares occasionally \citep{Paulson2006}. \citet{France2020} obtained X-ray (\emph{Chandra} ACIS-S) and FUV (\emph{HST} STIS/COS) data of this star both during quiescence and during a flare. We used these data to prepare both a quiescent DEM and a flare DEM, under the assumptions that the system is still in a collisionally dominated equilibrium and with solar coronal abundances scaled by the stellar metallicity [Fe/H] $= -0.32$. This is a very low S/N regime in the quiescent data but not nearly as low as TRAPPIST-1, discussed in Section \ref{sec:trappist_1}. A more physically accurate flare DEM would require adjusting the emissivity matrix to account for the magnetic reconnection's influence on the level populations. In the current framework, we see that the flaring state has more material than quiescence at temperatures between $\sim 10^5$ to $10^7$ K, but both the flare and quiescent DEMs agree beyond $3 \times 10^7$ K (see Figure \ref{fig:gj699_dem}). \citet{France2020} uses our quiescent and flaring model EUV spectra to investigate the influence of EUV variability on a hypothetical planet orbiting in the habitable zone of Barnard's star, demonstrating the applicability of our low-resolution spectra to models of atmospheric escape more complicated than simple energy-limited photoevaporation. We note that this DEM shape is very unusual compared to past published DEMs, but emphasize that the wide error intervals are likely to encompass the true DEM shape. If more data become available, perhaps the model constraints will narrow to something more familiar.

\subsection{TRAPPIST-1}\label{sec:trappist_1}
\begin{figure}
    \centering
    \includegraphics[width=0.5\textwidth]{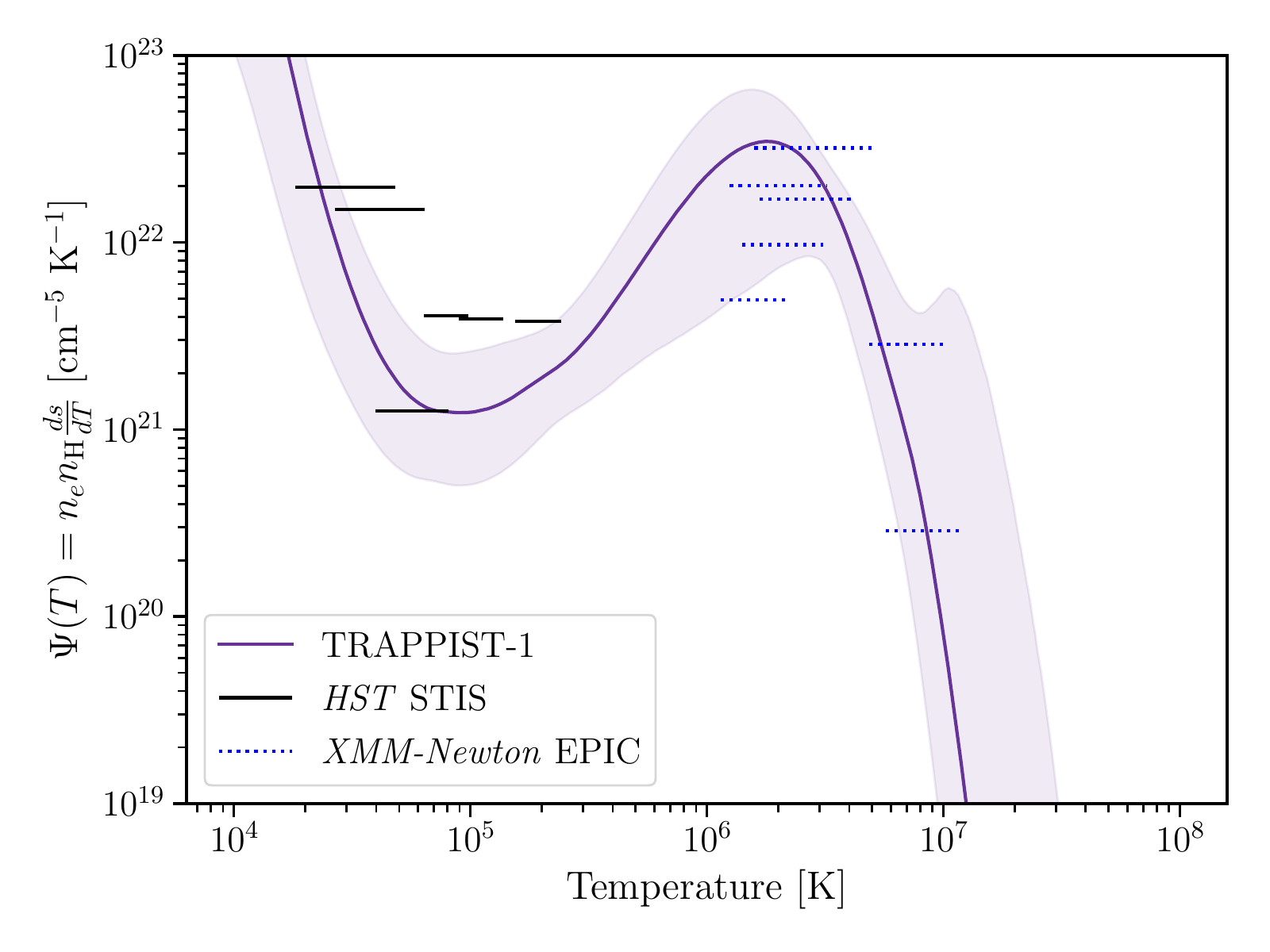}
    \caption{Comparing the DEM of TRAPPIST-1 fit to a quiescent X-ray spectrum from \emph{XMM-Newton} EPIC and line fluxes from STIS, with $\overline{\Psi}(T)$ constraints shown as horizontal dotted blue and solid black bars respectively.}
    \label{fig:trappist-one_dem}
\end{figure}

TRAPPIST-1 is an ultracool dwarf with a gaggle of seven planets discovered through transits \citep{Gillon2017}. It is included in the Mega-MUSCLES survey, an extension of the original MUSCLES survey from \citet{France2016}, and the measured FUV line fluxes (\emph{HST} STIS) and X-ray spectrum (\emph{XMM-Newton} EPIC) will be published shortly in Wilson et al. (submitted), where we discuss the implementation of our DEM method for this specific target. Wilson et al. (submitted) also compares the MegaMUSCLES spectral energy distribution of TRAPPIST-1 to the \texttt{PHOENIX} model published in \citet{Peacock2019a}. TRAPPIST-1 tests the fitting in a very low S/N regime constrained by a few FUV lines and a very faint X-ray spectrum, but we still get meaningful fits and constraints on the EUV flux (see Figures \ref{fig:trappist-one_dem} and \ref{fig:all_spectra}).

\subsection{Comparing the Entire Sample}
\begin{figure}
    \centering
    \includegraphics[width=\textwidth]{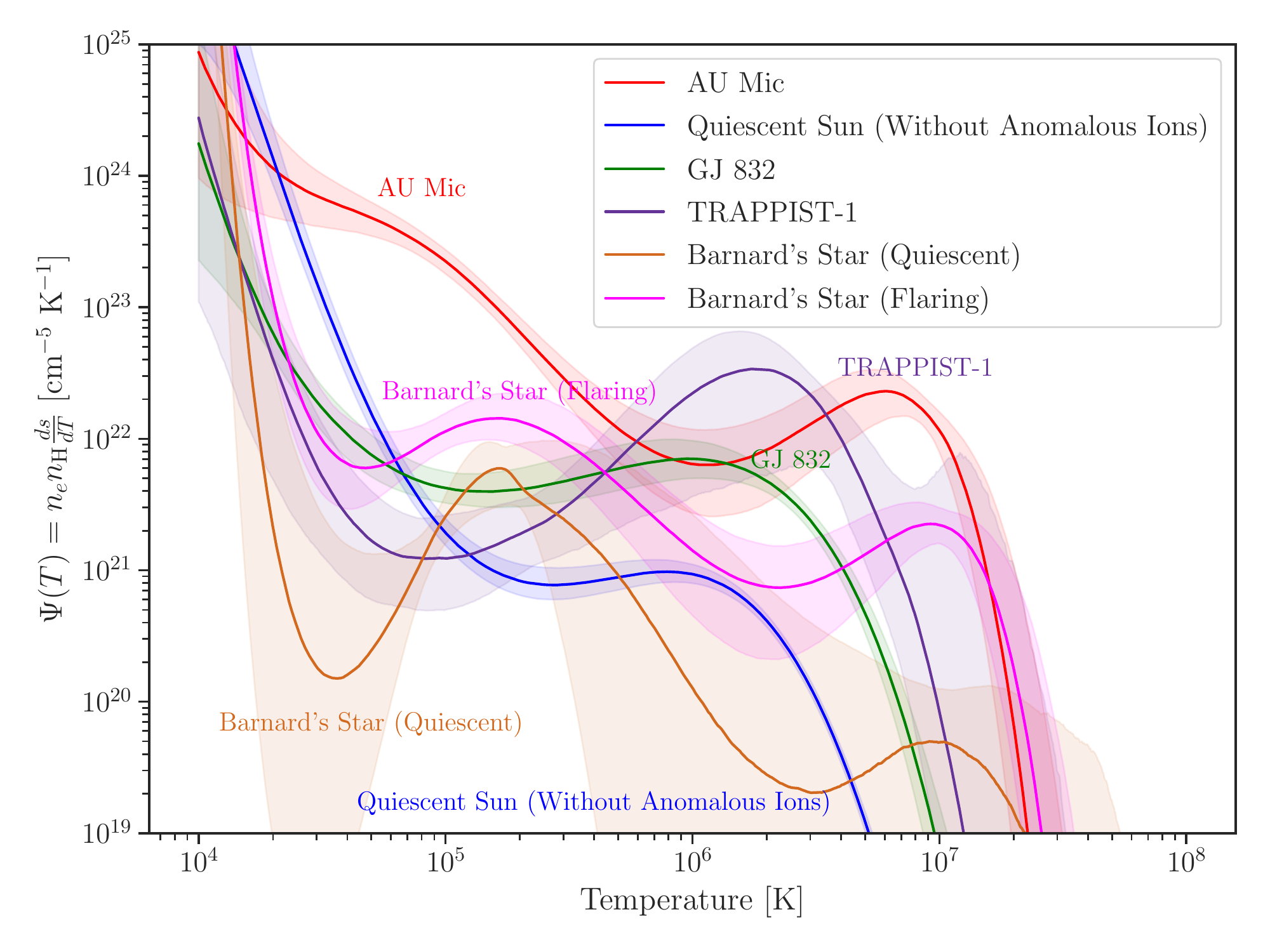}
    \caption{The DEMs of all stars considered in this work are plotted here with the solid line representing the median predicted value of $\Psi(T)$ and shaded intervals enclosing the 16$^{\textrm{th}}$ to 84$^{\textrm{th}}$ percentile intervals. Filling out this plot with more stars across a grid of effective temperature and stellar activity will allow us to investigate trends in the DEM and EUV spectra of cool dwarfs across the entire population.}
    \label{fig:all_dem}
\end{figure}

Figure \ref{fig:all_dem} shows the DEMs of all the stars considered within this work and shows some preliminary trends with activity and spectral type: more active stars have a higher mean DEM while the decline in the DEM associated with the move from the transition region to the corona appears to shift to higher temperatures at cooler spectral types. With a more complete sample, we could go a step further to interpolate the EUV flux of cool dwarfs that lack observed FUV and X-ray data by relating the DEM to the more accessible stellar parameters $T_{\textrm{eff}}$ and age, as traced by rotation and/or activity indicators from optical spectra. Determining the best method for this interpolation and testing its accuracy and precision is left to future work. 

\begin{figure}
    \centering
    \includegraphics[width=\textwidth]{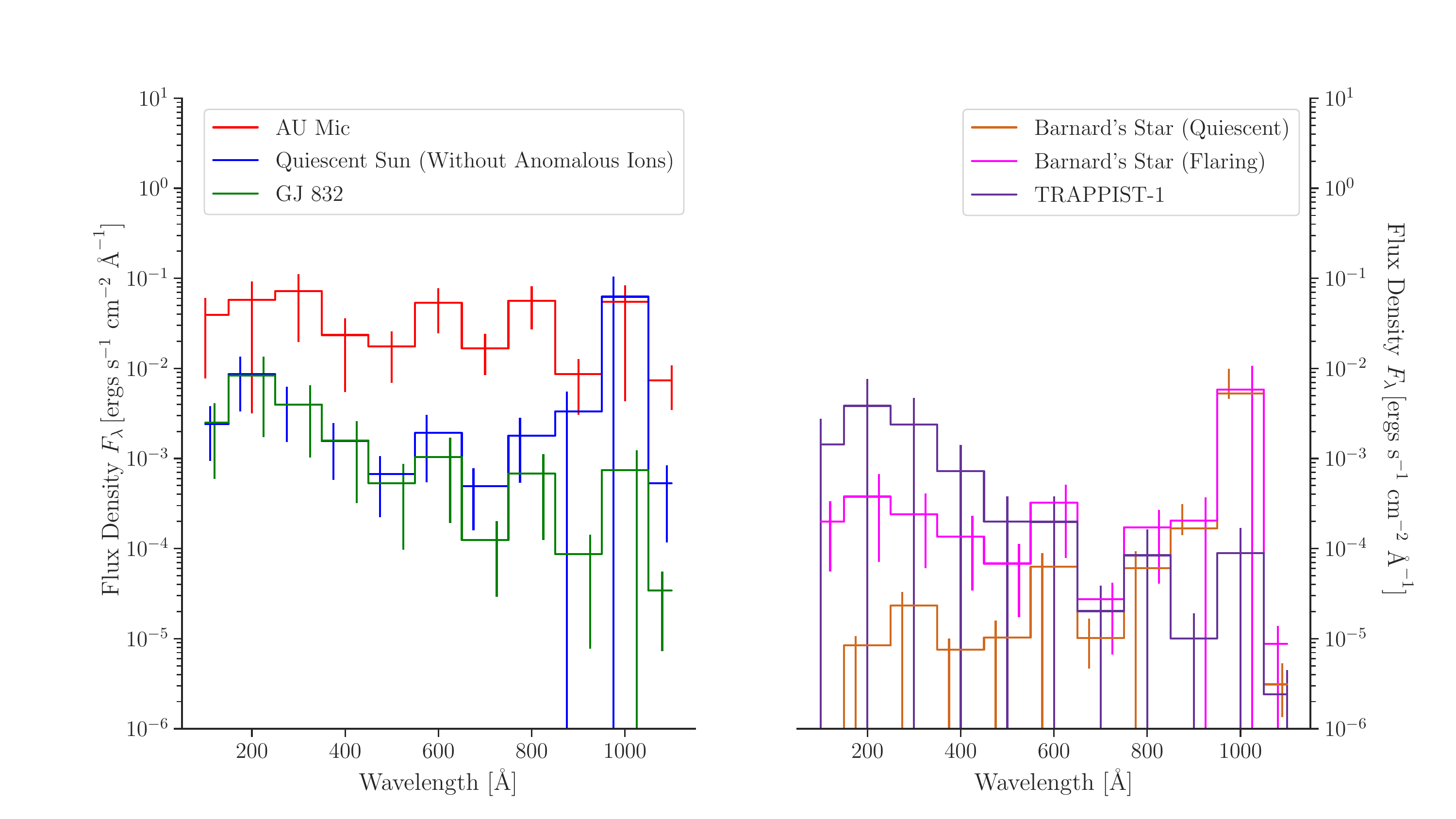}
    \caption{A comparison of all of our model spectra generated for the stars considered in this work, downsampled from a resolution of $R= \frac{\Delta \lambda}{\lambda}= 500$ to 100 $\textrm{\AA}$ wavelength bins, scaled to the flux received at 1 AU. The distance and radius assumed for each star is listed in Table \ref{table:all_euv_fluxes}, and can be scaled to surface flux or habitable zone distances by multiplying with the appropriate factor. On the left panel, we have the hotter stars of the sample: the Sun (spectrum generated from the DEM model without anomalous ions discussed in Section \S\ref{sec:euv_fitting}) in blue, AU Mic in red, and GJ 832 in green. On the right panel we have Barnard's Star in quiescence represented by the brown line, Barnard's Star while flaring shown in pink, and TRAPPIST-1 in dark purple. All the spectra shown have errorbars determined by the method described in Section \S\ref{sec:euv_fitting}.}
    \label{fig:all_spectra}
\end{figure}

We summarize the data products of this paper by plotting all of our DEM-generated EUV spectra scaled to the flux density observed at a distance of 1 AU from the host star in Figure \ref{fig:all_spectra}. The slope of the spectra across the EUV seems to vary as a function of both spectral type and activity. The hotter stars seem to have more EUV flux between 800 to 900 $\textrm{\AA}$ while the more active stars have more EUV flux between 100 to 600 $\textrm{\AA}$. These wavelength regions correspond to lines formed roughly at temperatures $2\times 10^5$ and $3\times 10^{6}$ K respectively (see Figure \ref{fig:gofnt_heatmap}). AU Mic, which is one of the more active stars and also one of the hotter stars in this sample, has a roughly flat EUV spectrum. TRAPPIST-1, which is also very active but much cooler, shows a strong negative slope from 100 to 1000 $\textrm{\AA}$. The spectral shape of the EUV is controlled by the relative strengths of the corona and chromosphere, and these preliminary observations of our EUV spectra conceptually agree with the findings of \citet{Linsky2020}. \citet{Linsky2020} measured the relationship between X-ray and Lyman-$\alpha$ flux for a large sample of FGKM dwarfs and found that for older and relatively inactive stars, the inverse relationship between coronal emission and effective temperature is much stronger than the inverse relationship between chromospheric emission and effective temperature. Trends in the shape of the EUV spectrum generated by the DEM should be investigated along with trends in the DEM and the data used to inform the fitting, and our current sample is simply too small to make stronger claims than these extremely tentative observations. Furthermore, the overall shape of the EUV spectrum may be significantly altered by the inclusion of continuum processes and an analysis of trends without this source of emissivity is premature. Table \ref{table:all_euv_fluxes} lists the integrated EUV flux at 1 AU, median $s-$factor, effective temperature, stellar radius, and distance for each star considered in this work.

\section{Conclusions and Future Work}\label{sec:conclusion}
Our tests with the Sun show that with a low-resolution X-ray spectrum and the strongest FUV emission lines, we can predict the EUV flux to within a factor of 2 across the entire EUV region. Furthermore, our characterization of the uncertainties in our method show that our predicted EUV spectra are consistent with the data for the Sun, with meaningful error bars to support that statement. We have demonstrated that our choice of functional form is able to describe the DEM within the temperature region relevant to predicting EUV flux, and that uncertainties in the abundance and average electron pressure can be accounted for and propagated to the final output spectra. While our approach to fitting the DEM has serious limitations discussed in Sections \S\ref{sec:euv_fitting} and \S\ref{sec:literature_comparison}, our tradeoff exchanging precision for simplicity allows us to fit DEMs and estimate EUV spectra for many more stars than more sophisticated methods which require EUV data or laborious iterations, both manual and computational. The method can be refined in the future to handle different stellar abundances more carefully, to use temperature-pressure profiles from stellar atmosphere models incorporating more physics, and to find a better way to handle the discrepant \texttt{CHIANTI} ionization equilibrium of the Na-like and Li-like isoelectronic sequences. Including the free-free and free-bound emission of hydrogen and helium species is possible with existing \texttt{CHIANTI} data and these emissivity sources will be accounted for in future DEM fits and EUV spectral reconstructions. The greatest problem with our current approach is the lack of elemental abundances tailored to individual stars based on their effective temperature and age. Updating the abundances we use and refitting the DEM may dramatically improve our precision, but this is left to future work.

\emph{Hubble} is the only observatory with FUV spectroscopy and the capability to observe a large sample of M dwarfs, but its lifetime is limited and there will be no replacement in the next few decades. Given the interest in M dwarf planetary systems, a number of survey programs have proposed using \emph{Hubble} to build up a spectral atlas of M dwarfs across the broad range of effective temperature and activity level represented within the spectral type. MUSCLES \citep{France2016}, HAZMAT \citep{Shkolnik2014}, FUMES (Pineda et al., in prep), and Mega-MUSCLES \citep{Froning2019} are completed programs with available data while observations for MEATS, a \emph{Hubble} survey targeting cool dwarf exoplanet hosts scheduled to be observed by \emph{Webb} (HST-GO-16166, PI-France), are forthcoming. Between all these surveys and archival data, we will have enough FUV and X-ray data of cool dwarfs to start characterizing them as a population: fitting DEMs to all nearby cool dwarf stars with sufficiently available data and estimating their EUV flux. With a sufficiently comprehensive DEM library, it may be possible to interpolate DEMs for stars too faint for FUV or X-ray observations, and build on the work of \citet{SanzForcada2011} to calculate the EUV luminosities of all known planet-hosting main-sequence stars. Until stellar atmosphere models become sophisticated enough to have a grid of models varying both effective temperature and stellar activity across the entire cool dwarf regime, or an EUV observatory \citep[e.g.]{France2019} is able to provide directly observed EUV spectra of nearby stars, differential emission measure techniques can satisfy the need for stellar EUV spectra which are physically informed and empirically calibrated.

\begin{splitdeluxetable*}{cccBcccc}
\tablehead{
\colhead{Star} & \colhead{Integrated 100 - 912 $\textrm{\AA}$ EUV Flux at 1 AU} & \colhead{Median $s$} & \colhead{Star} & \colhead{Effective Temperature} & \colhead{Stellar Radius} & \colhead{Distance}\\
\colhead{\NA} & \colhead{[erg s$^{-1}$ cm$^{-2}$]} & \colhead{\NA} & \colhead{\NA} & \colhead{[K]} & \colhead{[$R_\odot$]} & \colhead{[pc]}}
\startdata
\tablecaption{The DEM predictions for the integrated EUV flux at 1 AU and median $s-$factor uncertainty for each star considered in this work.\label{table:all_euv_fluxes}}
Sun (Woods et al. 2009) & $1.99$ & \NA & Sun (Woods et al. 2009) & 5772 \tablenotemark{[1]} &  1 & $4.848 \times 10^{-6}$ \\
Sun (DEM fit to FUV + X-ray)\tablenotemark{a} & $3.54^{+3.12}_{-2.12}$ & $0.60$ &  Sun (DEM fit to FUV + X-ray)\tablenotemark{a} & 5772\tablenotemark{[1]}&  1 & $4.848 \times 10^{-6}$\\
Sun (DEM fit without anomalous ions)\tablenotemark{b} & $1.99^{+1.28}_{-1.15}$ & $0.57$ & Sun (DEM fit without anomalous ions)\tablenotemark{b} & 5772\tablenotemark{[1]} & 1 & $4.848 \times 10^{-6}$\\
AU Mic & $31.9^{+21.8}_{-16.2}$  & $0.42$ & AU Mic & 3700\tablenotemark{[2]} & 0.75\tablenotemark{[2]} & 9.979\tablenotemark{[2]}\\
GJ 832 &  $1.66^{+1.30}_{-1.06}$ & $0.62$ &  GJ 832 &  3657\tablenotemark{[3]} & 0.499\tablenotemark{[4]} & 4.965\tablenotemark{[5]}\\
TRAPPIST-1 & $0.762^{+1.30}_{-0.744}$ & $0.74$ & TRAPPIST-1 & 2516\tablenotemark{[6]} & 0.121\tablenotemark{[6]}& 12.43\tablenotemark{[6]}\\
Barnard's Star (Quiescent) & $0.0183^{+0.109}_{-0.00860}$ & $0.43$ & Barnard's Star (Quiescent) &  3278\tablenotemark{[7]} & 0.178\tablenotemark{[7]} & 1.83\tablenotemark{[5]}\\
Barnard's Star (Flaring) & $0.146^{+0.112}_{-0.0965}$ & $0.45$ &  Barnard's Star (Flaring) & 3278\tablenotemark{[7]}&  0.178\tablenotemark{[7]} & 1.83\tablenotemark{[5]}\\
\enddata
\tablecomments{To enable scaling the spectrum to other quantities, we also list the stellar effective temperature, radius, and distance assumed in this work.}
\tablenotetext{a}{This integrated flux is from the DEM fit to the Sun using the X-ray spectrum and FUV lines.}
\tablenotetext{b}{This integrated flux is from the DEM fit to the Sun using the X-ray spectrum, EUV lines, and FUV lines excluding the anomalous ions \ion{N}{5}, \ion{C}{4}, and \ion{Si}{4}.}
\tablerefs{[1] \protect\citet{Mamajek2015}, [2] \protect\citet{Plavchan2020}, [3] \protect\citet{Bailey2009}, [4] \protect\citet{Houdebine2010}, [5] \protect\citet{Gaia2018}, [6] \protect\citet{VanGrootel2018}, [7] \protect\citet{Ribas2018}}
\end{splitdeluxetable*}

\acknowledgments

The authors thank the the referee for their constructive comments which improved the quality of this paper. G.M.D. acknowledges the contributions of conversations with Steve Cranmer in the initial stages of this project. This work was funded in part by grants HST-GO-14640, HST-GO-15071, and HST-GO-15264, which were provided by NASA through a grant from the Space Telescope Science Institute, which is operated by the Association of Universities for Research in Astronomy, Incorporated, under NASA contract NAS5-26555. ZKBT acknowledges funding for this work by the National Science Foundation under Grant No. 1945633. This research is based on observations made with the \emph{Far Ultraviolet Spectroscopic Explorer} and \emph{Extreme Ultraviolet Explorer}, obtained from the MAST data archive at the Space Telescope Science Institute, which is operated by the Association of Universities for Research in Astronomy, Inc., under NASA contract NAS 5–26555. We also used observations obtained with \emph{XMM-Newton}, an ESA science mission with instruments and contributions directly funded by ESA Member States and NASA. The analysis of these observations was supported by NASA/\emph{XMM-Newton} AO-17 grant number 82274, "A Unified Understanding of Flare Heating".

%

\vspace{5mm}
\facilities{HST(STIS, COS), FUSE, XMM-Newton, Chandra, EUVE.}


\software{astropy \citep{2013A&A...558A..33A},
CHIANTI \& ChiantiPy \citep{Dere1997, DelZanna2015},
emcee \citep{ForemanMackey2013},
matplotlib \citep{matplotlib},
numpy \citep{numpy}.}




\bibliography{sample63}{}
\bibliographystyle{aasjournal}



\end{document}